\documentclass[lettersize,journal]{IEEEtran}

\usepackage{amsmath,amsfonts}
\usepackage{amssymb}
\usepackage{mathtools}
\usepackage{amsthm}

\usepackage{algorithmic}
\usepackage{algorithm}

\usepackage{array}
\newcolumntype{C}[1]{>{\centering\arraybackslash}p{#1}}
\usepackage{booktabs}        
\usepackage{threeparttable}  
\usepackage{siunitx}         
\usepackage{multirow}
\usepackage{float}           

\usepackage[caption=false,font=normalsize,labelfont=sf,textfont=sf]{subfig}

\usepackage{textcomp}
\usepackage{stfloats}
\usepackage{url}
\usepackage{verbatim}
\usepackage{graphicx}
\usepackage{cite}
\usepackage{microtype}
\usepackage{hyperref}

\usepackage{enumitem}

\usepackage[capitalize,noabbrev]{cleveref}

\usepackage{pifont}
\newcommand{\cmark}{\ding{51}}
\newcommand{\xmark}{\ding{55}}
\newcommand{\pmark}{\(\sim\)}
\newcolumntype{C}[1]{>{\centering\arraybackslash}p{#1}}

\newcommand{\N}{\mathbb{N}}
\newcommand{\abs}[1]{\lvert#1\rvert}
\newcommand{\Pref}[2]{\mathrm{Pref}_{#1}\!\left(#2\right)}
\newcommand{\ispre}{\preceq}

\newcommand{\tightlist}{%
  \setlength{\topsep}{1pt}%
  \setlength{\itemsep}{0pt}%
  \setlength{\parsep}{0pt}%
  \setlength{\partopsep}{0pt}%
  \setlength{\leftmargin}{1.6em}%
}

\newcommand{\ABCstyle}{%
  \renewcommand{\theenumi}{\Alph{enumi}}%
  \renewcommand{\labelenumi}{(\theenumi)}%
}
\newcommand{\SLstyle}{%
  \renewcommand{\theenumi}{\Alph{enumi}}%
  \renewcommand{\labelenumi}{(\textsf{\theenumi})}%
}

\theoremstyle{plain}
\newtheorem{theorem}{Theorem}[section]

\newtheorem{lemma}[theorem]{Lemma}

\theoremstyle{definition}
\newtheorem{definition}[theorem]{Definition}
\newtheorem{assumption}[theorem]{Assumption}
\theoremstyle{remark}
\newtheorem{remark}[theorem]{Remark}

\usepackage[textsize=tiny]{todonotes}

\usepackage[most]{tcolorbox}

\newtcolorbox{highlightbox}[1][]{%
  colback=gray!15,
  colframe=black!60,
  boxrule=0.6pt,
  arc=6pt,
  left=10pt,right=10pt,top=8pt,bottom=8pt,
  enhanced,
  #1
}

\begin{document}

\title{Can Blockchains Reliably Train Machine Learning Models?}

\author{Peihao~Li,~%
\thanks{Peihao Li is with the Computer, Electrical and Mathematical Sciences and Engineering (CEMSE) Division, King Abdullah University of Science and Technology (KAUST), Thuwal, Saudi Arabia (e-mail: \href{mailto:peihao.li@kaust.edu.sa}{peihao.li@kaust.edu.sa}).}%
\and Nadia~Dahmani%
\thanks{Nadia Dahmani is with the Information Systems and Technology Management Department, College of Technological Innovation, Zayed University, Abu Dhabi, UAE (e-mail: \href{mailto:nadia.Dahmani@zu.ac.ae}{nadia.Dahmani@zu.ac.ae}).}%
}

\maketitle

\begin{abstract}
Large proof of work (PoW) networks allow anyone to earn rewards by running computation intensive hash puzzles for profit, yet they typically consume electricity comparable to that of medium sized countries. Repurposing computing resources from hash puzzles to machine learning training can benefit the energy sector as a whole, since this computing power is no longer wasted on solving hash puzzles but is instead used to train machine learning models that provide value across different application domains.

However, major technical gaps currently prevent this integration. To bridge these gaps, we introduce proof of training (PoT), a protocol that directs mining power toward verifiable training of machine learning models while preserving PoW’s incentives for participation and growth. We study PoT by theoretically identifying the blockchain structure that best meets the goals of training reliability, security, and scalability, and we further evaluate it by implementing a decentralized training network. Our results indicate considerable potential, including high task throughput, strong robustness, and improved network security. Our source code is available at: \url{https://github.com/Awesome-DeAI/proof-of-training}
\end{abstract}


\section{Introduction}

\IEEEPARstart{B}{lockchains} are distributed peer-to-peer (P2P) networks in which nodes collectively maintain a shared ledger by running a consensus protocol \cite{Bano2017SoKConsensus}.
Proof of work (PoW) is one of the most widely used consensus protocols and has been used for more than a decade \cite{Consensus1, Consensus2}. In PoW, network participants compete to be the first to solve a cryptographic puzzle and earn a reward. Since anyone can earn on a PoW blockchain by contributing resources, a PoW network’s computing power can scale dramatically as new participants join \cite{PoWhashrate_explode2, PoWhashrate_explode}, often growing by orders of magnitude within months. The apparent lack of a theoretical upper bound on the energy consumption of PoW networks such as Bitcoin and Ethereum has raised global concerns and has led to increasing governmental and institutional pressure as the industry attracts growing attention \cite{pressure2, prssure3, pressure1}. According to energy consumption analyses \cite{BTCConsumption2, BTCConsumption}, the yearly electricity consumption of the Bitcoin blockchain alone exceeds that of Sweden (131.79 TWh). This electricity and computing power are expended solely to enable \textit{trustless consensus}, without additional practical benefits. In 2022, Ethereum abandoned PoW, reducing power demand by 99.84\% to 99.99\% \cite{ETHPOW2POS, POWReduce}, but it also left substantial hashrate released, about 769.06 TH/s \cite{ETHHashRateAfterMerge}, with no specific application.

Meanwhile, as AI spreads across the economy and drives increasing demand for computation \cite{AIdemand}, efforts have been made to utilize the released computational resources for the machine learning sector. \cite{CoinAI} is an early proposal of proof of useful work PoUW, but it lacks customizable training tasks, which restricts its applicability to a limited set of business models. \cite{PoLe} addresses this by supporting client customized tasks, but its blockchain design has an inherent scalability flaw because storing test data in block bodies can quickly exhaust the storage capacity of consensus nodes. \cite{PoFL} innovatively recycles PoW energy into federated learning based training, and adds privacy preserving verification using homomorphic encryption. But the design assumes a trusted platform and reliable test data, and its privacy still depends on repeated interactive exchanges that can leak information over time and increase network overhead. \cite{PoDLEarliest} proposes proof of deep learning PoDL, a two phase design that relies on globally synchronized timestamps, making it fragile under network delay. PoDL also exposes the released test data to adversaries and requires frequent data exchanges that increase network overhead and reduce throughput. DLchain \cite{DLchain} improves on PoDL by removing reliance on a global clock and avoiding public exposure of test data, while adding mechanisms to prevent model theft. However, it requires near deterministic training reproducibility and therefore scales poorly to modern GPUs and non deterministic large model training. Similar issues are also found in distributed proof of deep learning D-PoDL \cite{D-PoLe}. \cite{collaborated_manager_learning} presents robust and efficient proof of learning RPoL, which inserts the block proposer address into the model to prevent model theft, but this introduces an address specific architectural change that may not hold under different tasks, optimizers, and training frameworks. RPoL is by default not multi model scalable because manager-side replay verification grows with the number of models being concurrently trained. In summary, current solutions still have major gaps in \textit{scalability} and \textit{security}, preventing their widespread adoption.

\begin{highlightbox}
\textbf{Why do scalability and security matter in a P2P learning network?}
\end{highlightbox}

\textit{Security} is important in blockchains because any adversary can join the network to cheat the system or steal funds from others. Recent surveys show that attacks target the network layer, the consensus and incentive layer, and the smart contract layer, and can compromise the integrity and availability of the global state \cite{Das2024LayerAttacks,Wen2021LayerSurvey,Wijewardhana2024PoWAttacks,Atzei2016EthereumSurvey}.Although attacks at the network layer and the smart contract layer can often be mitigated through software updates and improved programming practices \cite{network_attack_1, network_attack_2}, attacks that exploit protocol design flaws or vulnerabilities in the global state are far more challenging, and can sometimes be devastating \cite{consensus_bug_2, consensus_bug_1}. Distributed learning protocols without sufficient security considerations allow adversaries to claim rewards by faking reward mechanisms, which breaks fairness, reduces the value of trained models, and, most importantly, discourages participation.

\textit{Scalability} is equally important because it is a necessary condition for these protocols to make a \textit{real} industrial impact. Community developers contribute frequent upgrades after release and better tooling such as debuggers, testing support, and simulators, which in turn attract more contributors \cite{scalability_issue_1,scalability_issue_2}. Given the fast pace of the machine learning sector, if a protocol can handle only a few models at a time, it cannot create a wealth effect in the open market and will fail to attract developers and participants under winner-takes-most economic and network effects \cite{winnertake_1, winnertake_2, winnertake_3}, eventually becoming a dead chain. In addition, limited participation can also weaken security \cite{Croman2016Scaling}. Therefore, such protocols must keep verification and on chain state lightweight to scale, while still providing strong security.

In the rest of this paper, we present the proof of training (PoT) protocol design with the aim of bridging these technical gaps, and we analyze its \textit{security}, \textit{scalability}, and \textit{efficiency}. We further validate our analysis by implementing a decentralized training network and measuring throughput, robustness, and security.

\section{PoT Protocol Design}
\label{sec:pot_design}

\subsection{Consensus}

Blockchains are often distinguished by who can participate in consensus, with permissioned systems restricting it to known and controlled nodes managed by an organization or federation and permissionless systems allowing anyone to join and participate \cite{Bano2017SoKConsensus}. Considering that \textit{miners} in learning networks spend most of their computing resources on training tasks rather than on solving cryptographic puzzles as in PoW, we need to reconsider whether a PoW-like permissionless architecture is still feasible for PoT protocols. Since the PoT protocol includes an additional aggregation layer that aggregates canonical states, validates miners’ contributions and distributes rewards, we need to discuss whether network participants need permission to join this layer.

\begin{definition}
\label{def:pouw-model}
Time is discrete $t\in\N$. Each validator identity $i$ maintains a finite ledger (sequence) $L_i(t)$.
For sequences $A,B$, $A\ispre B$ denotes that $A$ is a prefix of $B$.
For a sequence $L$ and $\ell\in\{0,\dots,\abs{L}\}$, $\Pref{\ell}{L}$ is the prefix of $L$ of length $\ell$; $\abs{L}$ is the length of $L$.
We assume the following about the environment/protocol and the adversary $\mathcal{A}$:
\begin{enumerate}
\ABCstyle
\tightlist

\item \label{def:A}\textbf{(Network/adversary) Partitionable asynchrony.}
$\mathcal{A}$ schedules message delivery and may enforce a partition of validator identities into nonempty sets $P_1,P_2$
for an arbitrarily long duration such that \emph{no message sent by an identity in $P_1$ is delivered to any identity in $P_2$ and vice versa}
during that duration.
Within each $P_j$, messages sent between honest validators are eventually delivered.

\item \label{def:B}\textbf{(Protocol/model) PoT decoupling.}
ML-training participants do not send any consensus messages accepted by $\Pi$.
$\Pi$'s ledger progression depends only on validator-layer messages and record inputs.

\item \label{def:C}\textbf{(Adversary) Unresourced permissionless validators.}
For every $N\in\N$ and time $t$, $\mathcal{A}$ can activate $N$ fresh validator identities accepted by honest validators as valid participants,
without any scarce enforceable resource for admission/weight.

\item \label{def:D}\textbf{(Protocol/model) Mutual exclusion.}
There exist two valid records $x,y$ such that no valid finalized ledger is permitted to contain both $x$ and $y$.
\end{enumerate}
\end{definition}

\begin{definition}[Safety and liveness at confirmation depth $k$]
\label{def:pouw-props}
Fix $k\in\N$. A protocol $\Pi$ satisfies:
\begin{enumerate}
\SLstyle
\tightlist
\item \label{def:S}\textbf{$k$-persistence (safety).}
For all $t\le t'$ and honest identities $i$ at $t$, $j$ at $t'$, if $\abs{L_i(t)}\ge k$ then
\[
\Pref{\abs{L_i(t)}-k}{L_i(t)} \ispre L_j(t').
\]
\item \label{def:L}\textbf{Component-confirming liveness.}
For any partition component $P$ satisfying eventual internal delivery (per \ref{def:A}),
for any valid record $r$ submitted to an honest identity in $P$ at time $t_0$,
there exists $t\ge t_0$ such that for every honest $i\in P$,
\[
r \in \Pref{\abs{L_i(t)}-k}{L_i(t)}.
\]
\end{enumerate}
\end{definition}

\begin{theorem}[PoT: unresourced permissionless validators cannot achieve safety and liveness]
\label{thm:pouw}
Fix any $k\in\N$ and any ledger protocol $\Pi$ in the model of Def.~\ref{def:pouw-model}.
Under \ref{def:A}--\ref{def:D}, no protocol $\Pi$ can satisfy both \ref{def:S} and \ref{def:L} (Def.~\ref{def:pouw-props}).
\end{theorem}

\subsubsection{Notation for the proof}
Fix $k\in\N$. Recall that $L_i(t)$ denotes the (finite) ledger sequence held by identity $i$ at time $t$, and
$\abs{L_i(t)}$ is its length. For any sequence $L$ and $\ell\in\{0,\dots,\abs{L}\}$, $\Pref{\ell}{L}$ denotes the
prefix of $L$ of length $\ell$.

\smallskip
\noindent\textbf{Confirmed prefix (depth $k$).}
For any identity $i$ and time $t$, define the \emph{$k$-confirmed prefix} as
\[
\mathrm{Conf}_i(t)\;:=\;
\begin{cases}
\Pref{\abs{L_i(t)}-k}{L_i(t)}, & \text{if }\abs{L_i(t)}\ge k,\\[2pt]
\Pref{0}{L_i(t)}, & \text{if }\abs{L_i(t)}<k.
\end{cases}
\]
Intuitively, $\mathrm{Conf}_i(t)$ is the part of $L_i(t)$ that remains after discarding the last $k$ records.

\subsubsection{A tight Sybil lemma (identity-threshold confirmation is forgeable)}
\label{subsec:sybil-lemma}

The following lemma formalizes the fact that under unresourced permissionless admission (Def.~\ref{def:C}),
meaning admission is not tied to any scarce resource such as PoS stake,
any confirmation rule based solely on distinct identities can be satisfied via Sybil identities.

\begin{definition}[Identity-counting confirmation rule]
\label{def:id-pred}
Fix a ledger protocol $\Pi$ and a record $r$.
Consider any finite collection of protocol messages, each carrying a sender identity.
An \emph{identity-counting confirmation rule} is a function $\mathsf{Commit}_r$ that outputs $1$ exactly when, based only on what is written in those messages and what $\Pi$ can check locally from them, the messages provide enough evidence to treat $r$ as confirmed.

The rule may count how many \emph{distinct} sender identities support $r$ and may use any checks that $\Pi$ performs on a message, such as verifying signatures and verifying that a message is well formed.
The rule must not use any notion of external weight or admission cost, such as stake or a fixed membership roster.
\end{definition}

\begin{lemma}[Sybil inflation forces any identity-threshold confirmation]
\label{lem:sybil-forces-commit}
Assume Def.~\ref{def:C}. Fix any time interval during which a partition component $P$
has eventual internal delivery (Def.~\ref{def:A}). Fix any record $r$ and any honest identity
$h\in P$. If there exists \emph{any} finite set of syntactically valid messages $\mathcal{M}$
such that $\mathsf{Commit}_r(\mathcal{M})=1$ (Def.~\ref{def:id-pred}), then an adversary
$\mathcal{A}$ controlling only the identities it activates can, by activating sufficiently
many fresh identities inside $P$, cause $h$ to receive (eventually, within $P$) a set of messages
$\mathcal{M}'$ with $\mathsf{Commit}_r(\mathcal{M}')=1$ \emph{without} any cross-partition information.
\end{lemma}

\begin{proof}
By Def.~\ref{def:C}, for any $N$ the adversary can activate $N$ fresh validator identities whose
messages are accepted and processed as valid participants, with no scarce resources or weight.
Let $\mathcal{M}$ be a witness set with $\mathsf{Commit}_r(\mathcal{M})=1$.
The adversary activates enough fresh identities to supply distinct senders for all identities referenced
by $\mathcal{M}$, and has those identities emit the corresponding protocol messages (or any messages that
verify under the same checks used by $\mathsf{Commit}_r$). By Def.~\ref{def:A}, all such messages are eventually
delivered within $P$ to $h$. Since $\mathsf{Commit}_r$ depends only on locally checkable message and identity data,
$h$ will eventually hold a set $\mathcal{M}'$ satisfying $\mathsf{Commit}_r(\mathcal{M}')=1$.
\end{proof}

\begin{remark}
Lemma~\ref{lem:sybil-forces-commit} is not needed for the proof of Theorem~\ref{thm:pouw} given the liveness condition in Def.~\ref{def:L}.
We include it only to state explicitly a consequence of Def.~\ref{def:C}. If admission is not tied to any scarce resource, then any confirmation rule that relies only on counting distinct identities can be satisfied by creating enough new identities.
\end{remark}

\subsubsection{Proof of Theorem~\ref{thm:pouw}}
\label{subsec:proof_pouw_integrated}

\begin{proof}[Proof of Theorem~\ref{thm:pouw}]
Work in the model of Def.~\ref{def:pouw-model}. Let $x,y$ be valid records satisfying mutual exclusion
(Def.~\ref{def:pouw-model}). Choose two honest identities $h_1,h_2$ that are active from time $0$ onward.

\textbf{Step 1: create a sufficiently long partition.}
By Def.~\ref{def:A}, $\mathcal{A}$ can enforce a partition of the active identities into two nonempty
components $P_1,P_2$ with $h_1\in P_1$ and $h_2\in P_2$, with the following properties.
First, no messages are delivered across the cut between $P_1$ and $P_2$ for as long as $\mathcal{A}$ chooses.
Second, within each component there is eventual internal delivery.
In particular, $\mathcal{A}$ maintains such a partition from time $0$ onward and keeps it in place until it is
healed in Step~4.

\textbf{Step 2: inject mutually exclusive inputs.}
At time $t_0=0$, submit record $x$ to honest $h_1$ inside $P_1$ and submit record $y$ to honest $h_2$ inside
$P_2$. Both submissions are valid by assumption.

\textbf{Step 3: liveness forces conflicting confirmations inside components.}
While the partition remains in effect, apply component-confirming liveness \ref{def:L} to component $P_1$ and
record $x$. There exists a time $t_1\ge 0$ such that for every honest $i\in P_1$,
\[
x \in \mathrm{Conf}_i(t_1),
\]
and in particular $x \in \mathrm{Conf}_{h_1}(t_1)$.
Similarly, applying \ref{def:L} to component $P_2$ and record $y$ yields a time $t_2\ge 0$ such that
\[
y \in \mathrm{Conf}_{h_2}(t_2).
\]
Here we used exactly the statement of \ref{def:L} and no additional confirmation mechanism assumptions are
needed. Lemma~\ref{lem:sybil-forces-commit} explains why, under \ref{def:C}, any identity-threshold confirmation
rule would be forgeable inside each component.

\textbf{Step 4: heal the partition and derive a safety contradiction.}
Let $t^\star \ge \max\{t_1,t_2\}$ be a time when $\mathcal{A}$ stops withholding cross-cut messages and begins
delivering messages across $P_1,P_2$. Such a $t^\star$ exists because the cut in Step~1 can be maintained for an
arbitrarily long duration, and $\mathcal{A}$ may choose to heal it after both confirmations occur.
Now apply $k$-persistence \ref{def:S}.

Using \ref{def:S} with $i=h_1$ and $t=t_1$, and with $j=h_2$ and any $t'\ge t^\star$, gives
\[
\mathrm{Conf}_{h_1}(t_1)\ispre L_{h_2}(t') \quad \forall\, t'\ge t^\star,
\]
so $x$ must appear in all sufficiently late ledgers of $h_2$.

Using \ref{def:S} with $i=h_2$ and $t=t_2$, and with $j=h_1$ and any $t'\ge t^\star$, gives
\[
\mathrm{Conf}_{h_2}(t_2)\ispre L_{h_1}(t') \quad \forall\, t'\ge t^\star,
\]
so $y$ must appear in all sufficiently late ledgers of $h_1$.

Therefore, after $t^\star$, the stable ledger histories must contain both $x$ and $y$.
This contradicts mutual exclusion in Def.~\ref{def:pouw-model}, meaning no valid ledger can contain both $x$
and $y$. Hence \ref{def:S} and \ref{def:L} cannot simultaneously hold under \ref{def:A}--\ref{def:D}.
\end{proof}

\smallskip
\noindent\textbf{Design consequence.}
Under PoT decoupling \ref{def:B}, achieving both \ref{def:S} and \ref{def:L} requires violating \ref{def:C}:
validator admission or weight must be resource-bound (e.g., bonded/slashable stake or another scarce, enforceable resource).
For simplicity, we assume validators \textit{stake} at the validation layer to secure the protocol.

\subsection{Participants}

As solo miners become uncompetitive and mining power concentrates into large pools \cite{solo_converge_1, solo_converge_2, solo_converge_3}, we focus on a general protocol in which each miner submits a complete model for a given client task. We do not consider multiple miners training the same model, as this would substantially increase security and consensus complexity and is impractical, which is largely due to the rapid pace of machine learning with more efficient algorithms emerging daily. Overly specific protocols would require frequent upgrades, which is unrealistic because blockchain consensus updates are extremely difficult and require extensive coordination \cite{blockchain_upgrade_difficult1, blockchain_updgrade_difficult2}. Moreover, existing research on multi-node distributed training of the same model is flawed in terms of security and scalability to varying degrees. PoUW is similar to PoW in many financial aspects. We expect a successful training network to eventually consist of miners at large scale, which we call super miners, each coordinating many subnodes and distributing tasks for a given training job.

While miners are rewarded for providing computational resources and executing model training workloads, validators are rewarded for evaluating the models’ performance against the test sets clients provide, after which the validator subsequently broadcasts a separate validation message to the network for each model, including the model's performance and the identity of the miner, thereby quantifying the miner's contribution.

Aggregators are rewarded for preserving the integrity and consistency of the ledger’s canonical state and for executing the protocol rules honestly. Aggregators must stake a large volume of protocol tokens to be authorized as aggregator nodes, and such proof-of-stake (PoS) mechanisms have been extensively studied and proven secure \cite{pos_safe_1, pos_safe_2, pos_safe_3}.

We propose two types of participants in the consensus layer, with \textit{aggregator nodes} maintaining the ledger and \textit{validator nodes} evaluating models. This design leverages PoS, which has been proven secure for nearly a decade, while encouraging broader validator participation to strengthen security. Because validators require much less stake than aggregators, many participants can join to secure the training layer. Since aggregator nodes do not directly monitor the honesty of validators and miners, the consensus layer includes a hidden role called \textit{verifiers}, who challenge validators that behave adversarially. This role is hidden because any validator can act as a verifier when it detects wrongdoing in the network.

\subsection{Mechanisms}

Unlike other PoUW mechanisms in which every trained model is redundantly validated, which quickly consumes network bandwidth and creates a bottleneck, PoT follows the approach used in Ethereum Optimism rollups \cite{rollup_layer2, rollup_layer22}. In this approach, validators are assumed to be honest initially, but there is a challenge period during which anyone can dispute validator wrongdoing and receive a reward if the challenge succeeds. This frees aggregators from the burden of validating every model training task in the network, greatly improving scalability while maintaining network security through a proof dispute mechanism.

\subsubsection{PoT mechanisms and task cycle (integrated details)}
\label{subsec:pot_mechanisms_full}

Proof of Training (PoT) redirects distributed computation toward decentralized model training while preserving the open participation and incentive structure of blockchain systems. A client $\mathcal{C}$ posts a training task by publishing an initial model and training data. Service providers, also called miners, train candidate models and commit to their outputs before the validation data is released. Validators evaluate the revealed models with a deterministic scoring rule. Any network participant can act as a verifier by challenging incorrect validations during a dispute window. Aggregator nodes maintain a canonical global ledger $\mathcal{L}$, resolve disputes, and finalize rewards.

\paragraph{Participants and notation.}
We denote the set of $n$ aggregator nodes that maintain the global ledger by $\boldsymbol{\mathcal{G}}=\{\mathcal{G}_i\}_{i=1}^{n}$.
We use $\mathcal{C}$ for a client, $\mathcal{P}$ for a service provider, and $\mathcal{V}$ for a validator.
We use $\mathcal{N}$ for an arbitrary network participant, which may be a client, service provider, validator, or aggregator.
The network supports a verifier role through challenges, and this role is not tied to a dedicated identity type. Any $\mathcal{N}$ may submit a challenge as a verifier.

\paragraph{Keys and signatures.}
Each participant $\mathcal{N}$ holds participant specific security variables $\mathcal{S}_{\mathcal{N}}$ that include a public key $\mathcal{S}_{\mathcal{N}}[\textsc{pk}]$ and a private key $\mathcal{S}_{\mathcal{N}}[\textsc{sk}]$.
A signature on message $m$ is
\[
\sigma_{\mathcal{N}}(m) := \mathrm{Sig}_{\mathcal{S}_{\mathcal{N}}[\textsc{sk}]}(m),
\]
and verification is
\[
\mathrm{VRF\_Sig}\!\bigl(\mathcal{S}_{\mathcal{N}}[\textsc{pk}], \sigma, m\bigr)\in\{0,1\}.
\]

\paragraph{Models and identifiers.}
We write $\mathcal{M}$ for a model, and $\mathcal{M}_{\mathcal{C}}$ for the initial model posted by the client.
Training data and test data are denoted by $\mathcal{D}_{\text{train}}$ and $\mathcal{D}_{\text{test}}$.
To keep ledger messages compact, we represent a model by a fixed length identifier
\[
\mathrm{MID}(\mathcal{M}) := \mathrm{H}(\mathcal{M}),
\]
where $\mathrm{H}$ is a collision resistant hash function over a canonical serialization of $\mathcal{M}$.
When we write a signature on a model, we mean a signature on its identifier,
\[
\sigma_{\mathcal{P}}^{\mathcal{M}} := \mathrm{Sig}_{\mathcal{S}_{\mathcal{P}}}\!\bigl(\mathrm{MID}(\mathcal{M})\bigr).
\]
We use $\boldsymbol{\mathcal{M}}$ to denote a set of candidate models revealed for a task.

\paragraph{Deterministic scoring.}
Validators evaluate a revealed model using the protocol scoring function
\[
\mathrm{VRF\_Model}(\mathcal{M}, \mathcal{D}_{\text{test}})\rightarrow \text{score}.
\]
The protocol requires $\mathrm{VRF\_Model}$ to be deterministic across honest nodes. A client may select from a registry of supported evaluation rules, or provide parameters for a supported rule, but it does not supply arbitrary executable validation code. The scoring computation is intended to be lightweight relative to training, so that challengers can re evaluate claims when needed.

\paragraph{Remark on time in blockchain systems.}
PoT does not assume a globally synchronized wall clock. Phase boundaries are enforced by the ledger, typically by
block height. Concretely, a boundary can be specified by a reference height $H_0$ and a phase length $\Delta H$.
Given a target block interval $\Delta_{\text{blk}}$, the corresponding wall-clock duration is estimated as
$\widehat{\Delta T}\approx \Delta H\cdot \Delta_{\text{blk}}$.
For readability, we continue to write phase boundaries as times $t_0,t_1,\dots$ in what follows, with the
understanding that each $t$ refers to a ledger-enforced boundary.

\paragraph{Mechanism overview.}
A PoT task cycle is described by four polynomial time algorithms
\[
(\mathrm{Claim},\mathrm{Validate},\mathrm{Verify},\mathrm{Finalize}).
\]
Aggregators record claims and validations without re validating every model by default. Only disputed validations trigger additional work, which keeps the ledger footprint small when disputes are rare.

\paragraph{Claim and commit.}
A service provider commits to a trained model before the reveal deadline. This commitment prevents model theft after the validation data is released. Formally,
\[
\mathrm{PoT.Claim}\bigl(\mathcal{M}_{\mathcal{C}},\mathcal{D}_{\text{train}},\mathcal{S}_{\mathcal{P}}\bigr)
\rightarrow
\bigl(\sigma_{\mathcal{P}}^{\mathcal{M}}{}_{t_1}, \mathcal{M}_{t_2}\bigr),
\]
where $\sigma_{\mathcal{P}}^{\mathcal{M}}{}_{t_1}$ is the commitment published at time $t_1$ and $\mathcal{M}_{t_2}$ is the revealed model published at time $t_2$.
The phase boundaries satisfy
\begin{equation}
\label{TrainingTimeEq}
t_0<t_1<t_0+\Delta T_{\text{train}}\le t_2.
\end{equation}
After the commit deadline $t_0+\Delta T_{\text{train}}$, the ledger rejects new model commitments for the task. A revealed model is accepted only if its identifier matches the last committed identifier of the same provider and the commitment signature verifies.

\paragraph{Validation.}
At time $t_3$, the client releases $\mathcal{D}_{\text{test}}$ through a public link or content identifier recorded in $\mathcal{L}$.
Validators evaluate revealed models and broadcast signed validation reports.
Validation boundaries satisfy
\begin{equation}
\label{ValidationTimeEQ}
t_0+\Delta T_{\text{train}}\le t_3<t_4<t_5\le t_3+\Delta T_{\text{validate}}.
\end{equation}

\paragraph{Verification and disputes.}
Any participant $\mathcal{N}$ can check a validation report by recomputing $\mathrm{VRF\_Model}(\mathcal{M},\mathcal{D}_{\text{test}})$ and comparing it with the reported score.
If the report is incorrect, $\mathcal{N}$ can submit a signed challenge message. Challenge timing satisfies
\begin{equation}
\label{ChallengeTimeEq}
t_5\le t_6<t_7\le t_3+\Delta T_{\text{validate}}+\Delta T_{\text{Challenge}}.
\end{equation}

\paragraph{First successful challenger rule and processing fees.}
When multiple challenges target the same incorrect validation, only the first successful challenge is rewarded. The notion of first is defined by ledger order, meaning the smallest inclusion height or earliest finalization time in $\mathcal{L}$. Later challenges against the same validation do not receive rewards.

If a challenge against validator $\mathcal{V}$ succeeds, a fraction $\phi\in(0,1]$ of the validator stake $s_{\mathcal{V}}$ is slashed. Let $\gamma\in(0,1)$ be the challenger reward coefficient, with $\gamma=0.5$ as a typical setting. The first successful challenger receives $\gamma\phi s_{\mathcal{V}}$, and the remaining $(1-\gamma)\phi s_{\mathcal{V}}$ is treated as processing fees that are distributed to aggregators according to the ledger fee rule for the task. A successfully challenged validation is excluded from reward finalization.

\paragraph{Finalization and rewards.}
After the dispute window ends, aggregators finalize the task outcome. Let $\boldsymbol{\mathcal{M}}$ be the set of revealed models, and let $\boldsymbol{\pi}$ be the set of surviving validation reports after removing any reports invalidated by successful challenges. Aggregators select the best model by sorting models according to the deterministic score implied by the surviving reports. Ties are broken deterministically using a public rule, such as lexicographic order of model identifiers.

\subsubsection{Miscellaneous notes}

\begin{itemize}
    \item \textit{Security and cryptoeconomics.}
    Aggregators stake a significant amount of protocol value to participate in maintaining $\mathcal{L}$, and misbehavior results in slashing, which is consistent with the safety and liveness requirements discussed above. Validators also stake to prevent identity inflation and to support the dispute mechanism.
    When a validator report is successfully challenged, the first successful challenger receives a fraction $\gamma$ of the slashed stake, and the remainder is distributed to aggregators as processing fees. This rule aligns the incentive logic used below.

    \item \textit{Role composition.}
    A single node may act as a service provider, a validator, and a challenger, subject to the staking rules for each role. This improves resource utilization and supports broad monitoring participation. The protocol does not require a dedicated verifier population. It only requires that at least one rational party can profitably monitor when incorrect reports occur.

    \item \textit{Deterministic validation.}
    The dispute mechanism relies on the property that honest nodes obtain identical scores for the same model and test data under $\mathrm{VRF\_Model}$. For this reason, the protocol supplies the scoring rule and fixes implementation details that affect determinism, such as preprocessing, batching, and numeric settings. Clients choose among supported evaluation rules instead of supplying arbitrary validation code.

    \item \textit{Commitment scheme and model theft resistance.}
    The claim process implements a commit and reveal structure. During the commit period, a provider publishes a signature on the model identifier. During the reveal period, the provider publishes the model itself. Because the ledger rejects new commitments after the training phase ends, a malicious participant that observes a revealed model cannot retroactively create a valid earlier commitment for it without controlling the original provider key. This prevents straightforward model theft at reveal time.
\end{itemize}

\subsection{Security}

We first rule out aggregator compromise by requiring each aggregator $g\in\mathcal{G}$ to lock a stake $s_g\ge s_{\min}$ to participate in ledger operation, where $\mathcal{G}$ is the set of aggregators, $s_g$ is the stake of $g$, and $s_{\min}>0$ is the minimum stake. Let $S:=\sum_{g\in\mathcal{G}} s_g$ be the total aggregator stake and $S_{\mathcal{A}}:=\sum_{g\in\mathcal{G}_{\mathcal{A}}} s_g$ be the adversary-controlled stake, where $\mathcal{G}_{\mathcal{A}}\subseteq\mathcal{G}$ denotes the aggregators controlled by $\mathcal{A}$. Under a PBFT-style BFT protocol with stake-weighted voting power, safety and liveness hold if $S_{\mathcal{A}}<S/3$; compromising the system requires $S_{\mathcal{A}}\ge S/3$; and full control of block finalization requires $S_{\mathcal{A}}\ge 2S/3$ \cite{BFT2}. Even if an adversary could acquire enough stake to compromise the ledger and thereby make honest participants a minority, such an attack is often economically irrational. Major security incidents in cryptocurrency markets are frequently followed by price drawdowns of at least $25\%$ \cite{hack1, hack2}, and protocol level compromises tend to cause more than a $90\%$ price decline because they directly ruin confidence in the ledger itself \cite{luna_death}. Since an attacker must hold substantial governance stake to compromise the consensus layer, an attack collapses the token’s value and can trigger slashing, making it effectively suicidal, which implies that the consensus layer is strongly secure considering such attack costs.

Therefore our primary interest is whether the learning layer behaves reliably under honest and adversarial participants. In particular, we focus on the reliability of \textit{miner nodes} and \textit{validator nodes}, since they directly determine whether training outputs are correct and whether rewards are allocated fairly. To keep the notation consistent, we briefly summarize the key variables and economic parameters used in the learning layer before stating our assumptions and main security theorem.

Consider a single client task with reward $R>0$, which may be paid in stable coins or protocol tokens.
We parameterize the reward split by three coefficients $\alpha_G,\alpha_M,\alpha_V \in (0,1)$:
\begin{equation}
\alpha_G+\alpha_M+\alpha_V = 1.
\label{eq:split-sum}
\end{equation}
Here $\alpha_G R$ is paid to aggregator nodes as protocol fees,
$\alpha_M R$ is paid to the \emph{winning miner} whose model achieves the best performance score, and
$\alpha_V R$ is paid to validators. Generally we have
$\alpha_M \gg \alpha_V > \alpha_G$.

Each validator $\mathcal{V}_i$ locks a stake $s_{\mathcal{V}_i}\ge s_{\min}>0$ during the task cycle, where
$s_{\min}$ is a protocol minimum. Let $C$ denote the set of validators whose validation messages for this task are
\emph{not} successfully challenged during the challenge window, and let
$S_C:=\sum_{\mathcal{V}_j\in C} s_{\mathcal{V}_j}$.
To prevent the case where one participant spawns many validator identities, validators in $C$ share the validator
reward pool $\alpha_V R$ proportionally by stake:
\begin{equation}
r_{\mathcal{V}_i} \;=\; \alpha_V R \cdot \frac{s_{\mathcal{V}_i}}{S_C}\quad (\mathcal{V}_i\in C),
\qquad
r_{\mathcal{V}_i} = 0\quad (\mathcal{V}_i\notin C).
\label{eq:validator-reward}
\end{equation}

For a model $\mathcal{M}$ and test set $\mathcal{D}_{\text{test}}$ released at validation time,
validators compute a deterministic score using the protocol-provided scoring function
\begin{equation}
s(\mathcal{M}) \;:=\; \mathrm{VRF\_Model}(\mathcal{M},\mathcal{D}_{\text{test}})\in\mathbb{R}.
\label{eq:true-score}
\end{equation}
A validator $\mathcal{V}_i$ broadcasts a signed validation message that contains a reported score $\widehat{s}_{\mathcal{V}_i}(\mathcal{M})$. We call the message \emph{correct} if
\begin{equation}
\widehat{s}_{\mathcal{V}_i}(\mathcal{M}) = s(\mathcal{M}),
\label{eq:correct-validation}
\end{equation}
and \emph{incorrect} otherwise.

PoT follows a dispute design: validator reports are assumed correct unless successfully challenged.
Let $\Delta T_{\text{Challenge}}$ be the challenge window length.
If a challenge against validator $\mathcal{V}_i$ succeeds, validator $\mathcal{V}_i$ is slashed by a fraction $\phi\in(0,1]$ of its stake and excluded from $C$ for that task:
\begin{equation}
\text{successful challenge }
\;\Rightarrow\;
\text{$\mathcal{V}_i$ loses } \phi s_{\mathcal{V}_i}
\text{ and earns } 0.
\label{eq:slash}
\end{equation}
Let $\gamma\in(0,1)$ be the fraction of the slashed amount awarded to the successful challenger; then the challenger receives
\begin{equation}
\text{challenger reward } = \gamma\,\phi s_{\mathcal{V}_i}.
\label{eq:challenger-reward}
\end{equation}

Let $c_V>0$ be the cost for a validator to compute $s(\mathcal{M})$ and publish a correct report, and let
$c_V^{\mathrm{bad}}\in[0,c_V]$ be the cost of misreporting, which is often smaller since the validator may skip the computation.
We also consider external incentives outside the protocol reward mechanism. We upper bound the total
\emph{per-task} external gain to any single validator by $B_{\max}\ge 0$.

\begin{assumption}[Learning-layer game model]
\label{ass:game}
Consider a single task as a sequential game with stages:
\textsf{Claim} (miners commit), \textsf{Validate} (validators report),
\textsf{Challenge} (anyone challenges), and \textsf{Finalize} (rewards paid).
Assume:
\begin{enumerate}\ABCstyle\tightlist
  \item \label{ass:det}\textbf{Determinism.} The scoring function is deterministic as in \eqref{eq:true-score}.
  \item \label{ass:bind}\textbf{Binding commitment.} The commit--reveal used in $\mathrm{PoT.Claim}$ is binding: a miner cannot claim a model it did not commit to before the reveal deadline.
  \item \label{ass:detect}\textbf{Detectability.} Any incorrect validation message is successfully challenged within $\Delta T_{\text{Challenge}}$ with probability at least $q\in(0,1]$.
  \item \label{ass:bribe}\textbf{Bounded external gain.} Any extra gain from misreporting outside the protocol reward mechanism is bounded by $B_{\max}$ per task.
  \item \label{ass:cost}\textbf{Costs.} A validator pays $c_V$ to compute/report correctly; misreporting costs at most $c_V^{\mathrm{bad}}$. A challenger pays cost $c_{\mathrm{ch}}>0$ to verify and submit a challenge.
\end{enumerate}
\end{assumption}

\begin{definition}[Strictly dominant strategy]
\label{def:strictly-dominant}
Fix a validator $\mathcal{V}_i$. Let $A_i$ be its strategy set and let $a_{-i}$ denote the strategy profile of all
other participants. Let $u_i(a_i,a_{-i};\omega)$ be the realized payoff under randomness $\omega$.
A strategy $a_i^\star\in A_i$ is \emph{strictly dominant} if, for any $a_{-i}$ and any $a_i\in A_i\setminus\{a_i^\star\}$,
\begin{equation}
\mathbb{E}_{\omega}\!\left[u_i(a_i^\star,a_{-i};\omega)\right]
\;>\;
\mathbb{E}_{\omega}\!\left[u_i(a_i,a_{-i};\omega)\right].
\label{eq:strictly-dominant}
\end{equation}
\end{definition}

\begin{theorem}[Incentive security of validators and miners]
\label{thm:incentive-security}
Work under \eqref{eq:split-sum}--\eqref{eq:challenger-reward} and Assumption~\ref{ass:game}.
Suppose the slashing parameters satisfy, for every validator $\mathcal{V}_i$ with stake $s_{\mathcal{V}_i}$,
\begin{equation}
q\,\phi\, s_{\mathcal{V}_i} \;\ge\; B_{\max} + \bigl(c_V - c_V^{\mathrm{bad}}\bigr).
\label{eq:validator-ic}
\end{equation}
Then truthful validation, i.e., reporting \eqref{eq:true-score} so that \eqref{eq:correct-validation} holds,
is a strictly dominant strategy for validators.

Moreover, if challenger incentives satisfy
\begin{equation}
q\,\gamma\,\phi\, s_{\min} \;\ge\; c_{\mathrm{ch}},
\label{eq:challenger-ic}
\end{equation}
then there exists an equilibrium in which at least one rational party monitors and challenges, and therefore any
incorrect validation attempt survives the challenge window with probability at most $(1-q)$.
\end{theorem}

\subsubsection{Proof of Theorem~\ref{thm:incentive-security}}
\label{subsec:proof_incentive_integrated}

Before proving the theorem, we briefly explain the parameters that appear in the incentive conditions.
The cost $c_V>0$ is the resource cost for a validator to validate a model correctly, including obtaining
$\mathcal{M}$ and $\mathcal{D}_{\text{test}}$, computing the deterministic score
$s(\mathcal{M})=\mathrm{VRF\_Model}(\mathcal{M},\mathcal{D}_{\text{test}})$, and signing and broadcasting the
validation message. The misreporting cost $c_V^{\mathrm{bad}}\in[0,c_V]$ is typically smaller, since a validator that
misreports can skip the scoring computation and pay less compute and bandwidth. The quantity $B_{\max}\ge 0$ upper
bounds any extra per-task gain from misreporting outside the protocol reward mechanism. Finally, the detectability
parameter $q$ in Assumption~\ref{ass:detect} captures the probability that an incorrect validation is successfully
challenged within the challenge window. In practice, $q$ depends on protocol configuration and monitoring intensity;
aggregators can increase $q$ by jointly updating system parameters such as the challenge window length
$\Delta T_{\text{Challenge}}$, the monitoring incentives, and the verification rules that govern challenge acceptance.

\subsubsection*{Validators: truth-telling is strictly dominant}
Fix any validator $\mathcal{V}_i$ with stake $s_{\mathcal{V}_i}$, and consider a single model $\mathcal{M}$ in the
task. Let $H$ be the truthful strategy that computes $s(\mathcal{M})$ and broadcasts a report satisfying
\eqref{eq:correct-validation}. Let $B$ be any alternative strategy, including misreporting.

If $\mathcal{V}_i$ misreports and its report is incorrect, then by Assumption~\ref{ass:detect} it is successfully
challenged within $\Delta T_{\text{Challenge}}$ with probability at least $q$. In that event, $\mathcal{V}_i$ is
slashed by $\phi s_{\mathcal{V}_i}$ and earns $0$ by \eqref{eq:slash}. With complementary probability at most
$(1-q)$, the incorrect report survives and $\mathcal{V}_i$ may earn some validator reward. Since a single validator
cannot earn more than the full validator pool, we have $r_{\mathcal{V}_i}\le \alpha_V R$. Additionally, any extra
per-task gain from misreporting outside the protocol reward mechanism is bounded by $B_{\max}$
(Assumption~\ref{ass:bribe}). Therefore,
\begin{equation}
\mathbb{E}_{\omega}\!\left[u_i(B,a_{-i};\omega)\right]
\;\le\;
(1-q)\,\alpha_V R \;+\; B_{\max} \;-\; q\,\phi\, s_{\mathcal{V}_i} \;-\; c_V^{\mathrm{bad}}.
\label{eq:misreport-ub}
\end{equation}

Under $H$, the validator pays cost $c_V$ and its report is correct. By determinism (Assumption~\ref{ass:det}),
\eqref{eq:correct-validation} holds and a challenge against this report cannot succeed. Hence $\mathcal{V}_i$ is not
slashed and is not excluded from $C$. Regardless of the realized reward share, we have
\begin{equation}
\mathbb{E}_{\omega}\!\left[u_i(H,a_{-i};\omega)\right] \;\ge\; -c_V.
\label{eq:truth-lb}
\end{equation}

Combining \eqref{eq:misreport-ub} and \eqref{eq:truth-lb} yields
\begin{equation}
\begin{aligned}
\mathbb{E}_{\omega}\!\left[u_i(H,a_{-i};\omega)\right]
-
\mathbb{E}_{\omega}\!\left[u_i(B,a_{-i};\omega)\right]
\;&\ge\;
q\,\phi\, s_{\mathcal{V}_i} \;-\; B_{\max} \\
&\quad-\; \bigl(c_V-c_V^{\mathrm{bad}}\bigr).
\end{aligned}
\label{eq:gap}
\end{equation}
By \eqref{eq:validator-ic}, the right-hand side is nonnegative, and it is strictly positive if \eqref{eq:validator-ic}
holds strictly. Since $\mathcal{V}_i$ and $a_{-i}$ were arbitrary, truthful validation is a strictly dominant strategy
for validators by Definition~\ref{def:strictly-dominant}.

\subsubsection*{Challengers: monitoring can be sustained}
Consider an incorrect validation by some validator whose stake satisfies $s_{\mathcal{V}_i}\ge s_{\min}$.
A challenger who verifies and submits a challenge pays cost $c_{\mathrm{ch}}$ (Assumption~\ref{ass:cost}).
If the challenge succeeds (probability at least $q$ by Assumption~\ref{ass:detect}), the challenger receives
$\gamma\,\phi\, s_{\mathcal{V}_i}$ by \eqref{eq:challenger-reward}, which is at least $\gamma\,\phi\, s_{\min}$.
Therefore,
\begin{equation}
\mathbb{E}\!\left[u_{\mathrm{ch}}\right]
\;\ge\;
q\,\gamma\,\phi\, s_{\min} \;-\; c_{\mathrm{ch}}.
\label{eq:challenger-gap}
\end{equation}
By \eqref{eq:challenger-ic}, the right-hand side is nonnegative, so monitoring and challenging can be supported in
equilibrium. Consequently, any incorrect validation attempt survives the challenge window with probability at most
$(1-q)$.

\subsubsection*{Miners: best response is to maximize the true score}
Under dominant-strategy truthful validation, finalized validation reports match the deterministic score
$s(\mathcal{M})$ in \eqref{eq:true-score}, so miners cannot gain by manipulating validation outcomes. By binding
commitment (Assumption~\ref{ass:bind}), a miner cannot claim a model it did not commit to before the reveal deadline,
nor can it swap its committed model after seeing other reveals. Therefore a miner's winning probability depends only
on the true score of its committed model, and its best response is to maximize \eqref{eq:true-score} subject to its
training cost. This completes the proof.

\subsubsection*{Note on challenger rewards and $\gamma$}
A successful challenge must be validated by the aggregator nodes before rewards are finalized. This introduces extra
verification overhead at the ledger layer that scales with the aggregator set size, typically $\mathcal{O}(n)$ in the
number of aggregators. For this reason, the challenger reward fraction $\gamma$ in \eqref{eq:challenger-reward} is not
taken to be $1$ in practice. A typical setting is $\gamma\approx 0.5$, so that part of the slashed stake remains in
the protocol treasury or compensates aggregators for the extra work.

\subsection{Scalability}
\label{sec:agg-throughput}

We model the aggregator layer as a stake weighted BFT committee of size $n$ that finalizes the canonical ledger by repeatedly synchronizing a batch of pending transactions.

\begin{definition}[Throughput]
\label{def:tps}
Let $N_{\mathrm{tx}}(t_0,t_1)$ be the number of ledger transactions finalized in the time interval $[t_0,t_1]$.
The ledger throughput in transactions per second is
\begin{equation}
\mathrm{TPS}_{\mathrm{ledger}}
\;:=\;
\liminf_{T\to\infty}\frac{N_{\mathrm{tx}}(t,t+T)}{T}.
\label{eq:tps-def-ledger}
\end{equation}
If $N_{\mathrm{task}}(t_0,t_1)$ denotes the number of PoT tasks finalized in $[t_0,t_1]$, define the task throughput
\begin{equation}
\mathrm{TPS}_{\mathrm{task}}
\;:=\;
\liminf_{T\to\infty}\frac{N_{\mathrm{task}}(t,t+T)}{T}.
\label{eq:tps-def-task}
\end{equation}
\end{definition}

\begin{theorem}[Aggregator throughput bound]
\label{thm:agg-throughput}
Fix a committee size $n$. Let $b$ be the number of ledger transactions that the committee attempts to synchronize and finalize per consensus instance, and let $\delta$ be a representative one way network latency between honest aggregators.
Let $T_{\mathrm{cons}}(n,b,\delta)$ denote the steady state time needed to finalize one such batch under the committee protocol and network conditions.
Then the ledger transaction throughput satisfies
\begin{equation}
\mathrm{TPS}_{\mathrm{ledger}}
\;\le\;
\frac{b}{T_{\mathrm{cons}}(n,b,\delta)}.
\label{eq:tps-ledger-clean}
\end{equation}
If each PoT task induces at most $N_{\mathrm{tx,task}}$ ledger transactions, then the sustainable task throughput satisfies
\begin{equation}
\mathrm{TPS}_{\mathrm{task}}
\;\le\;
\frac{b}{T_{\mathrm{cons}}(n,b,\delta)\,N_{\mathrm{tx,task}}}.
\label{eq:tps-task-clean}
\end{equation}
\end{theorem}

\subsubsection{Aggregator throughput details}
\label{subsec:agg-throughput-details}

We bound the per batch consensus time by separating a latency component, a communication component, and a cryptographic verification component.

\paragraph{Latency bound.}
Let $\delta$ be a representative one way latency between honest aggregators.
Any BFT style committee protocol that requires a constant number of sequential message delivery rounds has a latency floor proportional to $\delta$.
We write this as
\begin{equation}
T_{\mathrm{cons}}(n,b,\delta) \;\ge\; c_{\mathrm{lat}}\,\delta,
\label{eq:tcons-lat-app}
\end{equation}
where $c_{\mathrm{lat}}>0$ is a protocol dependent constant.

\paragraph{Communication bound.}
Let $B_{\mathcal{G}}$ denote the sustainable consensus bandwidth of an aggregator in bytes per second.
Let $\mathrm{BytesPerBatch}(n,b)$ denote the total consensus traffic that must be delivered per finalized batch at committee size $n$ and batch size $b$.
Then
\begin{equation}
T_{\mathrm{cons}}(n,b,\delta) \;\ge\; \frac{\mathrm{BytesPerBatch}(n,b)}{B_{\mathcal{G}}}.
\label{eq:tcons-net-app}
\end{equation}
A standard abstraction for PBFT style voting and dissemination is
\begin{equation}
\mathrm{BytesPerBatch}(n,b)
\;=\;
b\,S_{\mathrm{tx}}
\;+\;
c_{\mathrm{msg}}\,n^{2}\,S_{\mathrm{vote}},
\label{eq:bytes-per-block-app}
\end{equation}
where $S_{\mathrm{tx}}$ is the mean transaction size, $S_{\mathrm{vote}}$ is the vote size, and $c_{\mathrm{msg}}>0$ captures the number of voting rounds and metadata.
The $n^{2}$ term reflects worst case all to all dissemination.
Certificate based designs can reduce the dominant communication growth by collecting votes at a leader and forming a compact certificate.

\paragraph{Computation bound.}
Let $\nu_{\mathrm{sig}}$ denote the number of signature verifications an aggregator can perform per second, and let $N_{\mathrm{sig}}(n)$ be the number of verifications required per finalized batch.
Then
\begin{equation}
T_{\mathrm{cons}}(n,b,\delta) \;\ge\; \frac{N_{\mathrm{sig}}(n)}{\nu_{\mathrm{sig}}}.
\label{eq:tcons-cpu-app}
\end{equation}
In certificate based workflows, $N_{\mathrm{sig}}(n)$ scales with the number of committee votes that must be checked to accept a certificate.

\paragraph{Combined bound.}
Since all components must complete before finalization, we have
\begin{equation}
T_{\mathrm{cons}}(n,b,\delta)
\;\ge\;
\max\!\left\{
c_{\mathrm{lat}}\,\delta,
\frac{\mathrm{BytesPerBatch}(n,b)}{B_{\mathcal{G}}},
\frac{N_{\mathrm{sig}}(n)}{\nu_{\mathrm{sig}}}
\right\}.
\label{eq:tcons-combined-app}
\end{equation}

\paragraph{Task accounting.}
A single task induces a bounded number of ledger transactions.
Order creation and finalization contribute $k_{\mathrm{task}}$, miners contribute $m$ claims, validators contribute $v$ reports, and disputes contribute an expected $\rho v$ challenge related transactions with $\rho\in[0,1]$.
Thus
\begin{equation}
N_{\mathrm{tx,task}}
\;=\;
k_{\mathrm{task}} \;+\; m \;+\; v \;+\; \rho v.
\label{eq:tx-per-task-app}
\end{equation}


\section{Protocol Implementation}

We implement PoT as a decentralized training network (DTN) that separates value settlement from utility execution.
The settlement layer anchors stake, escrow, and reward distribution on a mainstream chain, while the DTN execution
layer runs the task lifecycle, including order intake, model claiming, validation, dispute handling, and finalization.
This split keeps the value layer auditable while allowing the execution layer to evolve quickly as training workloads
and evaluation rules change. The public artifact repository \cite{pot_repo} provides the execution layer codebase, the
settlement contract interfaces, and the reproducibility package for the throughput and incentive experiments, and
serves as the canonical reference for message formats and system level details omitted here.

\begin{table}[b]
\centering
\scriptsize
\setlength{\tabcolsep}{1pt}
\renewcommand{\arraystretch}{1.10}
\begin{tabular}{p{0.48\linewidth} C{0.14\linewidth} C{0.18\linewidth} C{0.18\linewidth}}
\toprule
\textbf{Property} & \textbf{L1} & \textbf{L2} & \textbf{L1--L2} \\
\midrule
Final state committed on L1 & \cmark & \cmark & \cmark \\
Off chain heavy execution & \xmark & \cmark & \cmark \\
Utility transaction cost & High & Low & \textbf{0} \\
Value transaction cost & High & Low & Low \\
Throughput & Low & High & High \\
Bridge risk exposure & Low & High & Low \\
Failure isolation & Low & Medium & High \\
Scalability for utility workloads & \xmark & \cmark & \cmark \\
Upgradeability of core logic & Low & High & High \\
Operational complexity & Low & High & High \\
Development complexity & Low & High & High \\
\bottomrule
\end{tabular}
\caption{Compact comparison. \cmark = yes, \xmark = no, \pmark = depends.}
\label{tab:l1_l2_matrix}
\end{table}

\subsection{DTN architecture and implementation summary}

The DTN follows an L1--L2 design pattern similar in spirit to decentralized service networks \cite{chainlink2}.
On chain logic is intentionally minimal and stable, focusing on stake accounting, authorization, and settlement.
Most protocol operations are utility driven and occur off chain to avoid the cost, latency, and functional limits
of on chain execution \cite{tx_friction}. Concretely, clients post training orders, service providers commit and
reveal trained models, validators evaluate revealed models using deterministic scoring, and any participant may
challenge incorrect validation claims during a dispute window. An aggregator committee maintains a replicated global
ledger of task state and periodically commits finalized outcomes to the settlement contracts.

Many deployed blockchain and permissioned ledger systems finalize state with small committees, balancing fault tolerance,
coordination cost, and operational complexity. For example, EOSIO-style DPoS uses 21 elected block producers
\cite{eosio_consensus_doc,eosio_dpos_study}, BNB Smart Chain uses 21 validators per epoch \cite{bnb_bep131,bnb_security_case_study},
and TRON uses 27 super representatives \cite{tron_sr_doc,tron_committee_study}. In permissioned settings, IBFT-style BFT
deployments typically use single-digit to low-tens committees \cite{quorum_ibft_doc,ibft_analysis}, while Hyperledger Fabric
commonly uses a small ordering service cluster (e.g., 3, 5, 7, or 9 orderers) \cite{fabric_orderer_plan,fabric_arch_paper}.
Table~\ref{tab:committee_sizes} summarizes representative examples.

\begin{table}[b]
\caption{Examples of deployed systems and studies that discuss small committees that finalize ledger state.}
\label{tab:committee_sizes}
\centering
\footnotesize
\setlength{\tabcolsep}{4pt}
\renewcommand{\arraystretch}{1.15}
\begin{threeparttable}
\begin{tabular}{@{}p{0.36\columnwidth} p{0.44\columnwidth} p{0.12\columnwidth}@{}}
\toprule
\textbf{System} & \textbf{Management nodes} & \textbf{Size} \\
\midrule
EOSIO (DPoS) \cite{eosio_consensus_doc,eosio_dpos_study} &
Elected block producers that propose and finalize blocks &
21 \\
BNB Smart Chain \cite{bnb_bep131,bnb_security_case_study} &
Consensus validators that propose and finalize blocks &
21 \\
TRON (DPoS) \cite{tron_sr_doc,tron_committee_study} &
Super representatives that produce blocks &
27 \\
Permissioned IBFT deployments \cite{quorum_ibft_doc,ibft_analysis} &
BFT validator committee that finalizes blocks &
4--10 \\
Hyperledger Fabric \cite{fabric_orderer_plan,fabric_arch_paper} &
Ordering service cluster that establishes transaction order for commit &
3--9 \\
\bottomrule
\end{tabular}
\begin{tablenotes}[flushleft]
\footnotesize
\item \textit{Note:} ``Management nodes'' are nodes that finalize ledger state, either through block production and
finality voting or by ordering transactions for commit (Fabric). Delegated proof-of-stake deployments commonly use
fixed elected committees (e.g., 21 or 27). Permissioned BFT deployments often use single-digit to low-tens committees
for fault tolerance and operational simplicity.
\end{tablenotes}
\end{threeparttable}
\end{table}

\subsection{DTN Throughput Evaluation}
\label{sec:dtn_throughput}

DTN throughput reflects the scalability of PoT under realistic network constraints.
Aggregator nodes maintain the global ledger by synchronizing a transaction pool that includes orders, claims,
validations, and challenges. We evaluate synchronization time under a PBFT style workflow using the artifact
simulation that models committee size, link latency, and bandwidth caps \cite{BFT2}.

Many deployed blockchain and permissioned ledger systems finalize state with small committees, balancing fault tolerance,
coordination cost, and operational complexity. For example, EOSIO-style DPoS uses 21 elected block producers
\cite{eosio_consensus_doc,eosio_dpos_study}, BNB Smart Chain uses 21 validators per epoch \cite{bnb_bep131,bnb_security_case_study},
and TRON uses 27 super representatives \cite{tron_sr_doc,tron_committee_study}. In permissioned settings, IBFT-style BFT
deployments typically use single-digit to low-tens committees \cite{quorum_ibft_doc,ibft_analysis}, while Hyperledger Fabric
commonly uses a small ordering service cluster (e.g., 3, 5, 7, or 9 orderers) \cite{fabric_orderer_plan,fabric_arch_paper}.
Guided by these deployed ranges, we use committee sizes of 10, 30, and 50 as small, medium, and large configurations.

We consider small, medium, and large committees to reflect representative deployment settings, and we vary the number of
transactions synchronized per consensus instance. To translate protocol objects into a communication load, we use fixed-size
approximations consistent with the artifact implementation, where identifiers are SHA-256 hashes and signatures are RSA-2048
\cite{pot_repo}, and string fields contribute modest overhead. Under these assumptions, orders, validations, and challenges are
each on the order of a few hundred bytes per transaction in steady state, so batch size is the primary factor controlling
message volume.

\begin{table}[t]
\vspace{-2mm}
\caption{Synchronization time under different batch sizes, committee sizes, and bandwidth limits (s).}
\label{tab:syncTime2}
\centering
\scriptsize
\renewcommand{\arraystretch}{1.10}
\begin{threeparttable}
\begin{tabular*}{\columnwidth}{@{\extracolsep{\fill}}%
p{0.24\columnwidth}
p{0.26\columnwidth}
S[table-format=2.3]
S[table-format=1.3]
S[table-format=1.3]
@{}}
\toprule
\textbf{Scenario} & \textbf{Network size} & \multicolumn{3}{c}{\textbf{Bandwidth limit}} \\
\cmidrule(lr){3-5}
{\scriptsize (transactions per batch)} & {\scriptsize (committee nodes)} & \textbf{Slow} & \textbf{Medium} & \textbf{Fast} \\
\midrule
100 tx      & Small (10 nodes)  &  8.609 & 1.494 & 1.497 \\
100 tx      & Medium (30 nodes) &  8.707 & 1.685 & 1.755 \\
1{,}000 tx  & Medium (30 nodes) & 73.536 & 1.682 & 1.833 \\
100 tx      & Large (50 nodes)  &  8.697 & 1.842 & 1.752 \\
200 tx      & Large (50 nodes)  & 15.984 & 1.908 & 1.893 \\
5{,}000 tx  & Large (50 nodes)  & 37.532 & 1.767 & 1.678 \\
10{,}000 tx & Large (50 nodes)  & \multicolumn{1}{c}{--} & 7.215 & 2.074 \\
\bottomrule
\end{tabular*}
\begin{tablenotes}[flushleft]
\scriptsize
\item \textit{Note:} ``tx'' denotes transactions in a synchronization batch. ``--'' indicates no result available. Slow, Medium, and Fast cap each node's send and receive bandwidth at 0.1 Mbps, 30 Mbps, and 125 Mbps, respectively.
\end{tablenotes}
\end{threeparttable}
\end{table}

Table~\ref{tab:syncTime2} shows that bandwidth is the dominant practical limiter in globally distributed committees.
Under medium and fast bandwidth caps, synchronization time remains near a few seconds even as batch size grows,
whereas low bandwidth makes large batches impractical. Committee size also increases synchronization time through
voting and dissemination overhead, but its effect is substantially smaller than bandwidth in the regimes we study.
These results support a deployment configuration with tens of aggregators and sustained bandwidth on the order of
tens of Mbps per aggregator for robust high throughput operation.

Finally, the measured network throughput comfortably meets the scalability requirements implied by reported industrial job arrival rates. Public traces of deep learning clusters show average workload arrival rates well below one job per second when aggregated over multi-month periods \cite{model_demand_1,market_demand_2}. In this range, the DTN ledger provides sufficient capacity margin to support realistic training markets without ledger throughput becoming a limiting factor.

\subsection{DTN Learning Layer Security Simulations}
\label{sec:dtn_learning_security_sims}

This subsection instantiates the learning-layer game in Assumption~\ref{ass:game} and provides simulation evidence
consistent with the sufficient incentive conditions in Theorem~\ref{thm:incentive-security}.
In our model, learning-layer \textit{security} is an incentive property: validators prefer truthful reporting so that
finalized scores match the deterministic ground truth $s(\mathcal{M})$ in \eqref{eq:true-score}, and at least one
rational party prefers to monitor and challenge so that an incorrect validation survives the dispute window with
probability at most $(1-q)$ as stated in Theorem~\ref{thm:incentive-security}.
Exp1 visualizes the strict-dominance boundary implied by \eqref{eq:validator-ic}, Exp2 endogenizes detectability
through equilibrium monitoring consistent with \eqref{eq:challenger-ic}, and Exp3 links the resulting detectability
to repeated-task miner adaptation.

\subsubsection{Parameter settings}

Unless otherwise stated, we normalize the task reward to $R=1$ and set the minimum validator stake to $s_{\min}=1$.
For Exp1, we fix $s_{\mathcal{V}}=1$ and sweep detectability $q\in[0,1]$ together with the slashing exposure parameter
$\phi$, holding $B_{\max}=0.15$, $c_V=0.05$, and $c_V^{\mathrm{bad}}=0$; the payoff gap is evaluated on a
$251\times 251$ grid in $(q,\phi)$.
For Exp2, we model $n_{\mathrm{ch}}=200$ potential challengers with independent detection probability
$p_{\mathrm{detect}}=0.03$ and incorrect-attempt probability $p_{\mathrm{inc}}=0.15$, and we sweep the challenge cost
$c_{\mathrm{ch}}\in[0,0.15]$ over $16$ points; we use $\phi=0.8$, $\gamma=0.5$, $\tau=0.15$, and $\lambda=0.25$.
For Exp3, we simulate $n=60$ miners over $T=2500$ tasks with $\epsilon=0.08$ and $\eta=0.15$; high effort costs
$\Delta c=0.03$ and shifts the true-score mean from $\mu_{\mathrm{low}}=0$ to $\mu_{\mathrm{high}}=0.20$ with noise
$\sigma=0.35$.
We compare a high-detectability case using $q_{\mathrm{eff}}=q_{\mathrm{ref}}:=1-(1-p_{\mathrm{detect}})^{n_{\mathrm{ch}}}$
against a low-detectability case using $q_{\mathrm{eff}}=0.05$, and we report $95\%$ confidence bands over
$N_{\mathrm{seed}}=40$ independent random seeds.

\subsubsection{Exp1 validator dominance landscape}

Exp1 evaluates the validator payoff gap
\[
\Delta u
=\mathbb{E}\!\left[u_{\mathcal{V}}\!\left(\mathrm{truth}\right)\right]
-\mathbb{E}\!\left[u_{\mathcal{V}}\!\left(\mathrm{misreport}\right)\right],
\]
over a grid of $(q,\phi)$ and visualizes the boundary where truthful validation becomes strictly dominant.
Using the same per-task payoff comparison as in the proof of Theorem~\ref{thm:incentive-security}
(\S\ref{subsec:proof_incentive_integrated}), we instantiate
\begin{equation}
\label{eq:sim_exp1_gap}
\Delta u
=
q \phi s_{\mathcal{V}}
-
\left\{
B_{\max} + \left(c_V - c_V^{\mathrm{bad}}\right)
\right\}.
\end{equation}
Therefore, the indifference boundary $\Delta u=0$ is exactly the sufficient-condition threshold
\begin{equation}
\label{eq:sim_exp1_threshold}
q\phi s_{\mathcal{V}} = B_{\max} + (c_V-c_V^{\mathrm{bad}}),
\end{equation}
which matches \eqref{eq:validator-ic}.
In \cref{fig:dtn_learning_sims}, panel (a) overlays the numerically computed contour $\Delta u=0$ and the analytic
threshold \eqref{eq:sim_exp1_threshold}; the two coincide.
In the region where $\Delta u>0$, truthful validation strictly dominates misreporting for every validator, which is the
dominant-strategy statement in Theorem~\ref{thm:incentive-security}.

\subsubsection{Exp2 monitoring sustainability with endogenous detectability}

Exp2 endogenizes detectability by modeling monitoring as a participation equilibrium among $n_{\mathrm{ch}}$ potential
challengers. Let $n_{\mathrm{mon}}$ denote the number of challengers who choose to monitor.
If each monitoring challenger independently detects an incorrect validation with probability $p_{\mathrm{detect}}$, the
effective detectability becomes
\begin{equation}
\label{eq:sim_qeff}
q_{\mathrm{eff}} = 1-\left(1-p_{\mathrm{detect}}\right)^{n_{\mathrm{mon}}}.
\end{equation}
We approximate the \emph{first successful challenger} rule via mean-field reward dilution:
conditional on an incorrect validation occurring (probability $p_{\mathrm{inc}}$), a monitoring challenger detects it
with probability $q_{\mathrm{eff}}$ and is the first successful challenger with probability $\approx 1/n_{\mathrm{mon}}$.
Thus a representative monitoring challenger has expected payoff
\begin{equation}
\label{eq:sim_mon_payoff}
\pi_{\mathrm{mon}}
\approx
p_{\mathrm{inc}}\;q_{\mathrm{eff}}\;\frac{\gamma\,\phi\,s_{\min}}{n_{\mathrm{mon}}}
-
c_{\mathrm{ch}}.
\end{equation}
We map payoffs to participation using a soft best response with temperature $\tau$ and solve the induced fixed point
for $n_{\mathrm{mon}}$ using damping factor $\lambda$.

In \cref{fig:dtn_learning_sims}, panel (b) plots the equilibrium participation fraction $n_{\mathrm{mon}}/n_{\mathrm{ch}}$
and the induced $q_{\mathrm{eff}}$, together with the per-task incorrect-survival rate $p_{\mathrm{inc}}(1-q_{\mathrm{eff}})$.
The dashed vertical line marks the reference cost scale under full monitoring,
\[
c_{\mathrm{ref}}
:=
p_{\mathrm{inc}}\,q_{\mathrm{ref}}\,\gamma\,\phi\,s_{\min}/n_{\mathrm{ch}},
\qquad
q_{\mathrm{ref}}:=1-(1-p_{\mathrm{detect}})^{n_{\mathrm{ch}}},
\]
and the dotted line marks the empirical transition where the monitoring probability equals $0.5$ under the soft best
response. As $c_{\mathrm{ch}}$ increases, monitoring becomes less profitable, participation falls, and $q_{\mathrm{eff}}$
declines, increasing the survival rate of incorrect validations. This behavior is consistent with \eqref{eq:challenger-ic}:
monitoring can be sustained when expected slashing rewards cover verification costs.

\subsubsection{Exp3 repeated task miner adaptation under high and low detectability}

Exp3 links detectability to long-run miner incentives.
We simulate $n$ miners repeatedly competing for the miner reward share $\alpha_M R$ over $T$ tasks.
In each task, miner $i$ chooses effort $e_i\in\{0,1\}$, where $e_i=1$ incurs incremental cost $\Delta c$ but improves the
distribution of the miner's true score.
Conditional on effort, the true score is drawn as
\[
s_i \sim \mathcal{N}(\mu_{\mathrm{low}},\sigma^2)\ \text{ if } e_i=0,
\qquad
s_i \sim \mathcal{N}(\mu_{\mathrm{high}},\sigma^2)\ \text{ if } e_i=1.
\]
Winner selection is coupled to true score with probability $q_{\mathrm{eff}}$ and decoupled otherwise:
with probability $q_{\mathrm{eff}}$ the winner is $\arg\max_i s_i$, and with probability $(1-q_{\mathrm{eff}})$ the
winner is chosen uniformly at random. Under truthful validation (Theorem~\ref{thm:incentive-security}), higher
$q_{\mathrm{eff}}$ corresponds to stronger reward--quality coupling.

\begin{figure}[!t]
\centering
\subfloat[\textbf{Exp1.}]{%
  \includegraphics[width=0.48\columnwidth]{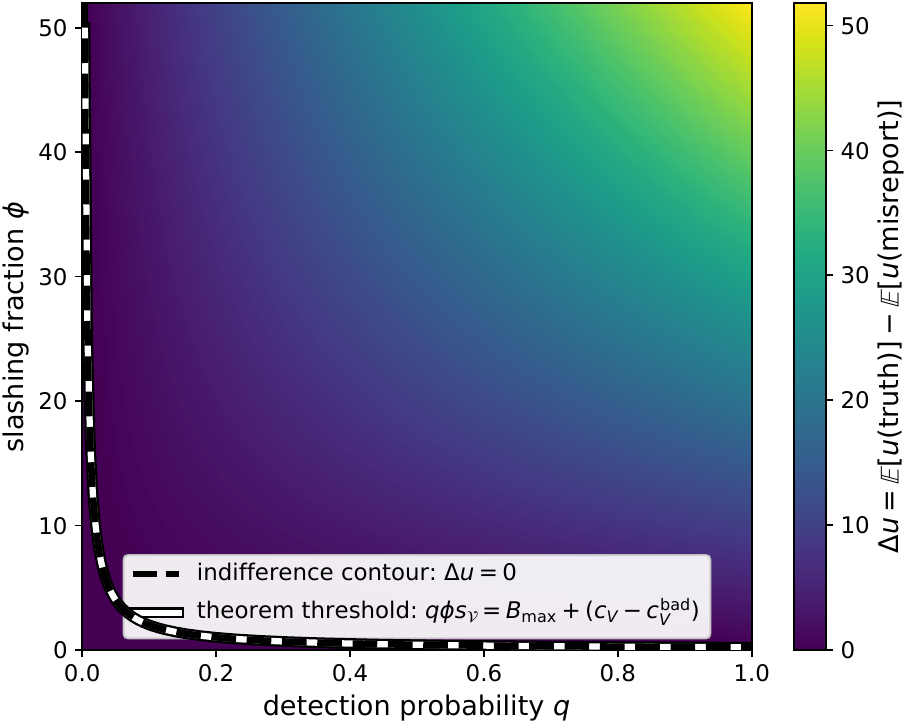}
}\hfill
\subfloat[\textbf{Exp2.}]{%
  \includegraphics[width=0.48\columnwidth]{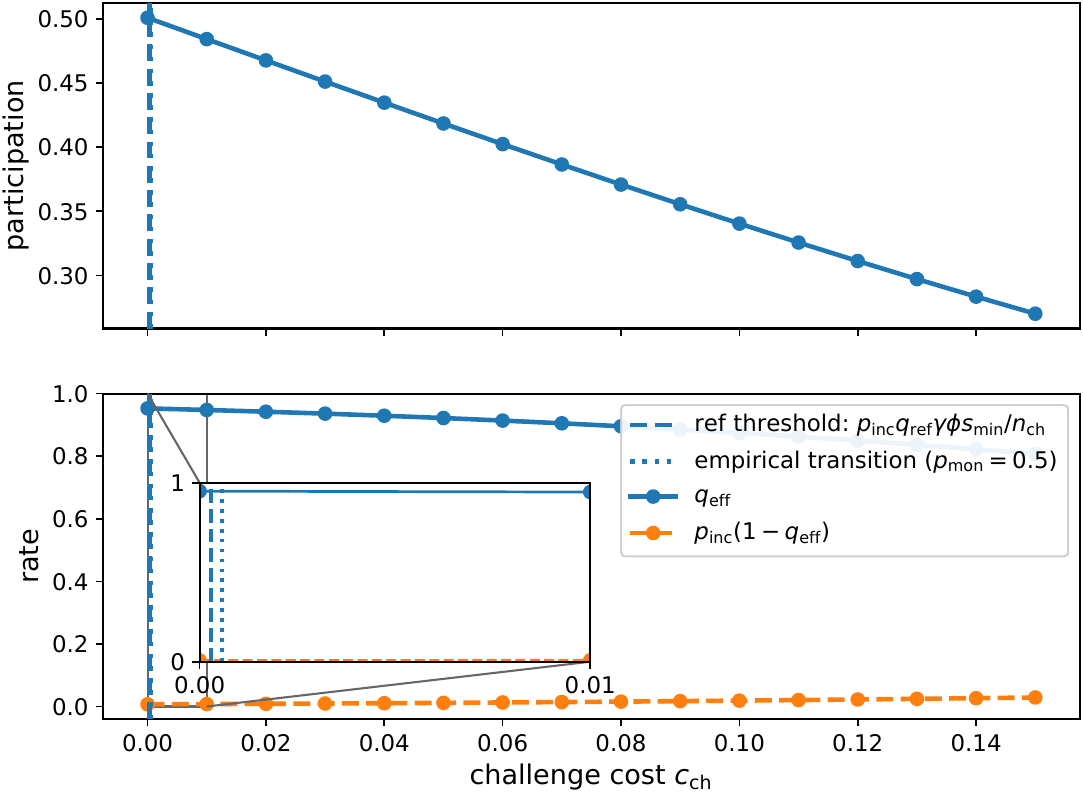}
}

\vspace{0.4em}

\subfloat[\textbf{Exp3 (high $q_{\mathrm{eff}}$).}]{%
  \includegraphics[width=0.48\columnwidth]{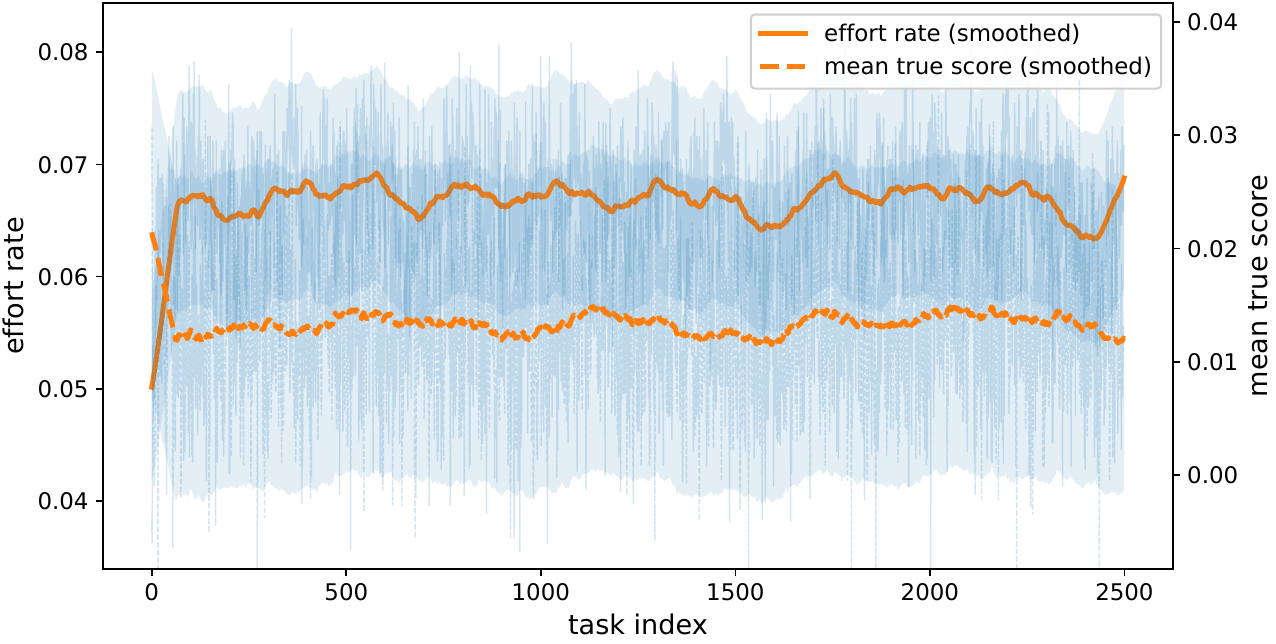}
}\hfill
\subfloat[\textbf{Exp3 (low $q_{\mathrm{eff}}$).}]{%
  \includegraphics[width=0.48\columnwidth]{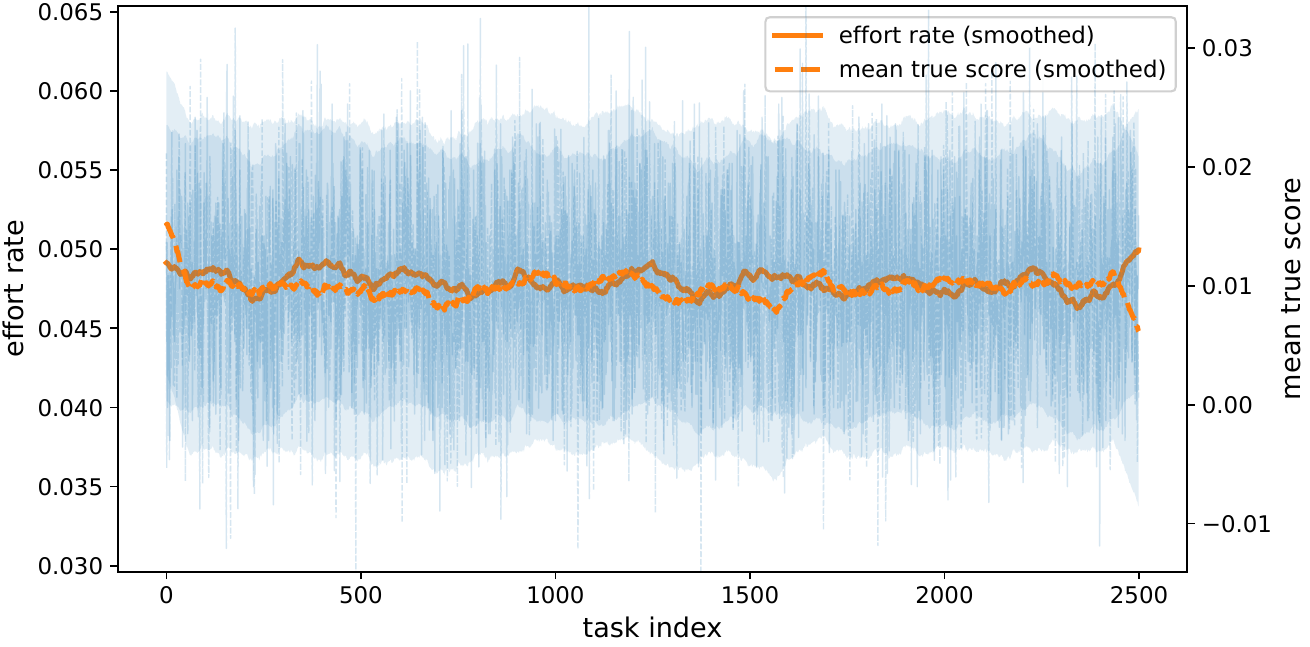}
}
\caption{\textbf{DTN learning layer security simulations.}
Exp1 plots the validator payoff gap $\Delta u$ over $(q,\phi)$ and overlays the indifference boundary $\Delta u=0$,
which coincides with the sufficient-condition threshold \eqref{eq:validator-ic}.
Exp2 endogenizes detectability through equilibrium monitoring and plots the induced $q_{\mathrm{eff}}$ together with
the incorrect-survival rate $p_{\mathrm{inc}}(1-q_{\mathrm{eff}})$ as challenge cost increases.
Exp3 links detectability to repeated-task adaptation: high detectability keeps rewards aligned with true score and
sustains effort, while low detectability decouples rewards from quality and reduces effort.}
\label{fig:dtn_learning_sims}
\end{figure}

In \cref{fig:dtn_learning_sims}, panel (c) reports the high-detectability case ($q_{\mathrm{eff}}=q_{\mathrm{ref}}$) and
panel (d) reports the low-detectability case ($q_{\mathrm{eff}}=0.05$).
Each panel plots the smoothed effort rate and smoothed mean true score over time, averaged over $N_{\mathrm{seed}}$
seeds with $95\%$ confidence bands.
Higher detectability sustains higher effort and higher realized quality, while low detectability reduces effort and
depresses true score. This illustrates the propagation mechanism underlying Theorem~\ref{thm:incentive-security}:
when validator truth-telling is dominant and monitoring sustains detectability, miners are incentivized to improve the
true score rather than to exploit validation noise.

\subsection{Additional discussion}

\subsubsection{System upgradability advantages}

Training workloads, evaluation rules, and tooling evolve rapidly, and a practical open training network must support
frequent updates without risking the safety of locked value. The L1--L2 split in Table~\ref{tab:l1_l2_matrix} supports
this by keeping stake and settlement anchored on chain while allowing the execution layer to evolve off chain through
standard secure update pipelines \cite{off_chain_upgrade_pattern1}. Major protocol changes that require coordination
can be executed through multi party authorization and proxy style upgrade patterns \cite{offchain_multisig,off_chain_upgrade_pattern2_1},
while governance sensitive monetary parameters can be gated by on chain voting and execution rules \cite{Dao_voting_contract,Dao_voting_contract_2}.
This separation improves failure isolation and allows emergency patches to the execution layer without exposing the
treasury or stake to execution layer faults.

\subsubsection{Scalable training interfaces}

PoT exposes a single miner single model interface at settlement, but it does not constrain how computation is
organized off chain. In practice, a miner can be a large cluster operator or a pool coordinator that aggregates many
contributors, mirroring the coordination patterns observed in proof of work mining pools \cite{solo_converge_1,solo_converge_3}.
If a miner coordinates internal workers $U_m$ that contribute measurable work units $w_u$, then a simple internal
payout rule is
\begin{equation}
p_u \;=\; \alpha_M R \cdot \frac{w_u}{\sum_{v\in U_m} w_v},
\qquad u\in U_m,
\end{equation}
which preserves the protocol interface while allowing diverse off chain accounting policies.

The same separation applies to large model training and privacy sensitive collaborative training. A miner can run
standard data parallel and model parallel workflows for large models \cite{megatron_lm,zero_rajbhandari2020,gpipe}
and still reveal a single final model for deterministic evaluation, or it can coordinate encrypted federated learning
using secure aggregation and related primitives \cite{fedavg2017,secureagg_ccs2017,ckks2017,abadi2016dp}. In all cases,
validators score the revealed model using the protocol supplied deterministic rule, and the dispute mechanism applies
without changing the settlement interface.
\section{Conclusion}

This paper investigates how to redirect computing resources from proof of work networks toward training machine learning models while preserving the security and scalability guarantees required for open participation and mass adoption. We justify our design choices with theoretical and empirical evidence, clarifying the participation and incentive structures needed for reliable operation in an open setting. PoT aligns rewards with true training quality by making honest reporting profitable and cheating economically unattractive, while keeping verification lightweight through a dispute mechanism. It scales by pushing heavy training and validation off-chain and recording only compact outcomes for settlement, allowing many tasks to run concurrently without congesting the ledger. Finally, we highlight that the protocol supports a broad range of workloads, including large model training and privacy sensitive collaborative training, without changing the basic interface, positioning PoT as a practical foundation for scalable and adversarially robust open training networks.

\bibliographystyle{IEEEtran}
\bibliography{example_paper}

@article{Consensus1,
author = {Nguyen, Truong and Kim, Kyungbaek},
year = {2018},
month = {01},
pages = {101-128},
title = {A survey about consensus algorithms used in Blockchain},
volume = {14},
journal = {Journal of Information Processing Systems},
doi = {10.3745/JIPS.01.0024}
}

@article{Consensus2,
author = {Wang, Wenbo and Dinh Thai, Hoang and Hu, Peizhao and Xiong, Zehui and Niyato, Dusit and Wang, Ping and Wen, Yonggang and Kim, Dong In},
year = {2019},
month = {01},
pages = {1-1},
title = {A Survey on Consensus Mechanisms and Mining Strategy Management in Blockchain Networks},
volume = {PP},
journal = {IEEE Access},
doi = {10.1109/ACCESS.2019.2896108}
}

@article{BTCConsumption,
author = {Alshahrani, Hani and Islam, Noman and Syed, Darakhshan and Sulaiman, Adel and Al Reshan, Mana and Rajab, Khairan and Shaikh, Asadullah and Shuja-Uddin, Jaweed and Soomro, Aadar},
year = {2023},
month = {02},
pages = {1510},
title = {Sustainability in Blockchain: A Systematic Literature Review on Scalability and Power Consumption Issues},
volume = {16},
journal = {Energies},
doi = {10.3390/en16031510}
}

@misc{BTCConsumption2,
  author = {Boyle, Jack},
  title = {{BTC Mining Used More Electricity than Sweden}},
  year = {2021},
  howpublished = {\url{https://beincrypto.com/btc-mining-used-more-electricity-than-sweden/}},
  note = {Accessed: [Date Accessed]}
}

@article{pressure1,
  author  = {Finney, Bradley R.},
  title   = {Win-Win Environmental Regulations for Crypto Mining: Developing a Regulatory Program That Reduces Environmental Harm and Promotes Innovation and Competition},
  journal = {Boston College Law Review},
  year    = {2024},
  volume  = {65},
  number  = {4},
  pages   = {1185--1250},
  month   = apr,
  note    = {Earlier version posted Aug. 28, 2023. Available at SSRN (No. 4554735).},
  url     = {https://ssrn.com/abstract=4554735}
}

@techreport{pressure2,
  author      = {Stoll, Christian and Klaa{\ss}en, Lena and Gallersd{\"o}rfer, Ulrich and Neum{\"u}ller, Alexander},
  title       = {Climate Impacts of Bitcoin Mining in the U.S.},
  institution = {MIT Center for Energy and Environmental Policy Research (CEEPR)},
  year        = {2023},
  month       = jun,
  number      = {CEEPR WP 2023-11},
  type        = {Working Paper},
  url         = {https://ceepr.mit.edu/wp-content/uploads/2023/06/MIT-CEEPR-WP-2023-11.pdf}
}

@article{prssure3,
title = {Cryptocurrency regulation and market quality},
journal = {Journal of International Financial Markets, Institutions and Money},
volume = {84},
pages = {101744},
year = {2023},
issn = {1042-4431},
doi = {https://doi.org/10.1016/j.intfin.2023.101744},
url = {https://www.sciencedirect.com/science/article/pii/S1042443123000124},
author = {Todd Griffith and Danjue Clancey-Shang}
}

@article{ETHPOW2POS,
  author  = {Asif, Rameez and Hassan, S. R.},
  title   = {Shaping the future of Ethereum: exploring energy consumption in Proof-of-Work and Proof-of-Stake consensus},
  journal = {Frontiers in Blockchain},
  volume  = {6},
  year    = {2023},
  doi     = {10.3389/fbloc.2023.1151724},
  url     = {https://www.frontiersin.org/journals/blockchain/articles/10.3389/fbloc.2023.1151724/full}
}

@article{POWReduce,
title = {Cryptocurrencies on the road to sustainability: Ethereum paving the way for Bitcoin},
journal = {Patterns},
volume = {4},
number = {1},
pages = {100633},
year = {2023},
issn = {2666-3899},
doi = {https://doi.org/10.1016/j.patter.2022.100633},
url = {https://www.sciencedirect.com/science/article/pii/S2666389922002653},
author = {Alex {De Vries}}
}

@misc{ETHHashRateAfterMerge,
  author        = {Kiffer, Lucianna and Skorik, Sophia and Vonlanthen, Yann and Wattenhofer, Roger},
  title         = {The PoW Landscape in the Aftermath of The Merge},
  year          = {2023},
  eprint        = {2310.01028},
  archivePrefix = {arXiv},
  primaryClass  = {cs.CE},
  url           = {https://arxiv.org/abs/2310.01028}
}

@article{AIdemand,
  title={Sustainable AI: Environmental Implications, Challenges and Opportunities},
  author={Carole-Jean Wu and Ramya Raghavendra and Udit Gupta and Bilge Acun and Newsha Ardalani and Kiwan Maeng and Gloria Chang and Fiona Aga Behram and James Huang and Charles Bai and Michael K. Gschwind and Anurag Gupta and Myle Ott and Anastasia Melnikov and Salvatore Candido and David Brooks and Geeta Chauhan and Benjamin Lee and Hsien-Hsin S. Lee and Bugra Akyildiz and Maximilian Balandat and Joe Spisak and Ravi Kumar Jain and Michael G. Rabbat and Kim M. Hazelwood},
  journal={ArXiv},
  year={2021},
  volume={abs/2111.00364},
  url={https://api.semanticscholar.org/CorpusID:240354766}
}

@inproceedings{PoDLEarliest,
   title={Energy-recycling Blockchain with Proof-of-Deep-Learning},
   url={http://dx.doi.org/10.1109/BLOC.2019.8751419},
   DOI={10.1109/bloc.2019.8751419},
   booktitle={2019 IEEE International Conference on Blockchain and Cryptocurrency (ICBC)},
   publisher={IEEE},
   author={Chenli, Changhao and Li, Boyang and Shi, Yiyu and Jung, Taeho},
   year={2019},
   month=may, pages={19–23} }

@InProceedings{DLchain,
author="Chenli, Changhao
and Li, Boyang
and Jung, Taeho",
editor="Ferreira, Joao Eduardo
and Palanisamy, Balaji
and Ye, Kejiang
and Kantamneni, Siva
and Zhang, Liang-Jie",
title="DLchain: Blockchain with Deep Learning as Proof-of-Useful-Work",
booktitle="Services -- SERVICES 2020",
year="2020",
publisher="Springer International Publishing",
address="Cham",
pages="43--60",
isbn="978-3-030-59595-1"
}

@INPROCEEDINGS{collaborated_manager_learning,
  author={Zhang, Xiaoli and Xu, Zhicheng and Cheng, Hongbing and Che, Tong and Xu, Ke and Wang, Weiqiang and Zhao, Wenbiao and Wang, Chunping and Li, Qi},
  booktitle={2023 IEEE 43rd International Conference on Distributed Computing Systems (ICDCS)}, 
  title={Secure Collaborative Learning in Mining Pool via Robust and Efficient Verification}, 
  year={2023},
  volume={},
  number={},
  pages={794-805},
  doi={10.1109/ICDCS57875.2023.00012}}

@article{PoFL,
author = {Qu, Xidi and Wang, Shengling and Hu, Qin and Cheng, Xiuzhen},
title = {Proof of Federated Learning: A Novel Energy-Recycling Consensus Algorithm},
year = {2021},
issue_date = {Aug. 2021},
publisher = {IEEE Press},
volume = {32},
number = {8},
issn = {1045-9219},
url = {https://doi.org/10.1109/TPDS.2021.3056773},
doi = {10.1109/TPDS.2021.3056773},
journal = {IEEE Trans. Parallel Distrib. Syst.},
month = aug,
pages = {2074–2085},
numpages = {12}
}

@InProceedings{D-PoLe,
author="Su, Xiangyu
and Larangeira, Mario
and Tanaka, Keisuke",
editor="Li, Shujun
and Manulis, Mark
and Miyaji, Atsuko",
title="Provably Secure Blockchain Protocols from Distributed Proof-of-Deep-Learning",
booktitle="Network and System Security",
year="2023",
publisher="Springer Nature Switzerland",
address="Cham",
pages="114--136",
isbn="978-3-031-39828-5"
}

@inproceedings{PoLe,
  author    = {Lan, Yixiao and Liu, Yuan and Li, Boyang and Miao, Chunyan},
  title     = {Proof of Learning (PoLe): Empowering Machine Learning with Consensus Building on Blockchains (Demo)},
  booktitle = {Proceedings of the {AAAI} Conference on Artificial Intelligence},
  year      = {2021},
  volume    = {35},
  number    = {18},
  pages     = {16063--16066},
  doi       = {10.1609/aaai.v35i18.18013},
  url       = {https://doi.org/10.1609/aaai.v35i18.18013}
}

@Article{CoinAI,
AUTHOR = {Baldominos, Alejandro and Saez, Yago},
TITLE = {Coin.AI: A Proof-of-Useful-Work Scheme for Blockchain-Based Distributed Deep Learning},
JOURNAL = {Entropy},
VOLUME = {21},
YEAR = {2019},
NUMBER = {8},
ARTICLE-NUMBER = {723},
URL = {https://www.mdpi.com/1099-4300/21/8/723},
PubMedID = {33267437},
ISSN = {1099-4300},
DOI = {10.3390/e21080723}
}

@article{PoWhashrate_explode,
title = {Valuing cryptocurrencies: A model of price and hashrate},
journal = {Finance Research Letters},
volume = {86},
pages = {108346},
year = {2025},
issn = {1544-6123},
doi = {https://doi.org/10.1016/j.frl.2025.108346},
url = {https://www.sciencedirect.com/science/article/pii/S1544612325016009},
author = {William C. Johnson}
}

@article{PoWhashrate_explode2,
  author       = {Kim, Daehan and Ryu, Doojin and Webb, Robert I.},
  title        = {Does a higher hashrate strengthen Bitcoin network security?},
  journal      = {Financial Innovation},
  volume       = {10},
  pages        = {79},
  year         = {2024},
  doi          = {10.1186/s40854-023-00599-8},
  url          = {https://doi.org/10.1186/s40854-023-00599-8},
  received     = {2023-02-22},
  accepted     = {2023-12-27},
  published    = {2024-06-05}
}

@inproceedings{Bano2017SoKConsensus,
author = {Bano, Shehar and Sonnino, Alberto and Al-Bassam, Mustafa and Azouvi, Sarah and McCorry, Patrick and Meiklejohn, Sarah and Danezis, George},
title = {SoK: Consensus in the Age of Blockchains},
year = {2019},
isbn = {9781450367325},
publisher = {Association for Computing Machinery},
address = {New York, NY, USA},
url = {https://doi.org/10.1145/3318041.3355458},
doi = {10.1145/3318041.3355458},
booktitle = {Proceedings of the 1st ACM Conference on Advances in Financial Technologies},
pages = {183–198},
numpages = {16},
location = {Zurich, Switzerland},
series = {AFT '19}
}

@misc{Das2024LayerAttacks,
  author = {Joydip Das and Syed Ashraf Al Tasin and Md. Forhad Rabbi and Md Sadek Ferdous},
  title  = {Analysing Attacks on Blockchain Systems in a Layer based Approach},
  note   = {arXiv:2409.10109},
  year   = {2024}
}

@article{Wen2021LayerSurvey,
  author  = {Yujuan Wen and Fengyuan Lu and Yufei Liu and Xinli Huang},
  title   = {Attacks and countermeasures on blockchains: A survey from layering perspective},
  journal = {Computer Networks},
  volume  = {191},
  pages   = {107978},
  year    = {2021}
}

@misc{Wijewardhana2024PoWAttacks,
  author = {Dinitha Wijewardhana and Sugandima Vidanagamachchi and Nalin Arachchilage},
  title  = {Examining Attacks on Consensus and Incentive Systems in Proof of Work Blockchains: A Systematic Literature Review},
  note   = {arXiv:2411.00349},
  year   = {2024}
}

@inproceedings{Croman2016Scaling,
  author    = {Kyle Croman and Christian Decker and Ittay Eyal and Adem Efe Gencer and Ari Juels and Ahmed E. Kosba and Andrew Miller and Prateek Saxena and Elaine Shi and Emin Gun Sirer and Dawn Song and Roger Wattenhofer},
  title     = {On Scaling Decentralized Blockchains},
  booktitle = {Financial Cryptography and Data Security Workshops},
  pages     = {106 to 125},
  year      = {2016}
}

@misc{Atzei2016EthereumSurvey,
  author = {Nicola Atzei and Massimo Bartoletti and Tiziana Cimoli},
  title  = {A survey of attacks on Ethereum smart contracts},
  note   = {IACR Cryptology ePrint Archive, Report 2016/1007},
  year   = {2016}
}

@inproceedings {network_attack_1,
author = {Ethan Heilman and Alison Kendler and Aviv Zohar and Sharon Goldberg},
title = {Eclipse Attacks on {Bitcoin{\textquoteright}s} {Peer-to-Peer} Network},
booktitle = {24th USENIX Security Symposium (USENIX Security 15)},
year = {2015},
isbn = {978-1-939133-11-3},
address = {Washington, D.C.},
pages = {129--144},
url = {https://www.usenix.org/conference/usenixsecurity15/technical-sessions/presentation/heilman},
publisher = {USENIX Association},
month = aug
}

@article{network_attack_2,
title = {A comprehensive survey of smart contract security: State of the art and research directions},
journal = {Journal of Network and Computer Applications},
volume = {226},
pages = {103882},
year = {2024},
issn = {1084-8045},
doi = {https://doi.org/10.1016/j.jnca.2024.103882},
url = {https://www.sciencedirect.com/science/article/pii/S1084804524000596},
author = {Guangfu Wu and HaiPing Wang and Xin Lai and Mengmeng Wang and Daojing He and Sammy Chan}
}

@article{consensus_bug_1,
title = {Formal verification of fraud-resilience in a crowdsourcing consensus protocol},
journal = {Computers \& Security},
volume = {131},
pages = {103290},
year = {2023},
issn = {0167-4048},
doi = {https://doi.org/10.1016/j.cose.2023.103290},
url = {https://www.sciencedirect.com/science/article/pii/S0167404823002006},
author = {Hamra Afzaal and Muhammad Imran and Muhammad Umar Janjua}
}

@misc{consensus_bug_2,
      title={ZKPs: Does This Make The Cut? Recent Advances and Success of Zero-Knowledge Security Protocols}, 
      author={Stavros Kassaras and Leandros Maglaras},
      year={2020},
      eprint={2006.09990},
      archivePrefix={arXiv},
      primaryClass={cs.CR},
      url={https://arxiv.org/abs/2006.09990}, 
}

@article{scalability_issue_1,
  author    = {Bosu, Amiangshu and Iqbal, Ahsan and Shahriyar, Rifat and Chakraborty, Pritom and Iqbal, Anindya and Rahman, A. S. M.},
  title     = {Understanding the motivations, challenges and needs of Blockchain software developers: a survey},
  journal   = {Empirical Software Engineering},
  volume    = {24},
  pages     = {2636--2673},
  year      = {2019},
  doi       = {10.1007/s10664-019-09708-7},
  url       = {https://doi.org/10.1007/s10664-019-09708-7},
  note      = {Published 27 April 2019; issue date 15 August 2019}
}

@inproceedings{scalability_issue_2,
  author    = {Voloder, Ante and di Angelo, Monika},
  title     = {Comparison of Smart Contract Platforms from the Perspective of Developers},
  booktitle = {Blockchain -- ICBC 2023},
  editor    = {Wang, Qiang and Feng, Jian and Zhang, L.-J.},
  series    = {Lecture Notes in Computer Science},
  volume    = {14206},
  publisher = {Springer},
  address   = {Cham},
  year      = {2023},
  doi       = {10.1007/978-3-031-44920-8_7},
  url       = {https://doi.org/10.1007/978-3-031-44920-8_7},
  isbn      = {978-3-031-44920-8},
  note      = {Published 01 October 2023}
}

@article{winnertake_1,
title = {Blockchain competition: The tradeoff between platform stability and efficiency},
journal = {European Journal of Operational Research},
volume = {296},
number = {3},
pages = {1084-1097},
year = {2022},
issn = {0377-2217},
doi = {https://doi.org/10.1016/j.ejor.2021.05.031},
url = {https://www.sciencedirect.com/science/article/pii/S0377221721004598},
author = {Shangrong Jiang and Yuze Li and Shouyang Wang and Lin Zhao}
}

@article{winnertake_2,
  author    = {Aufiero, Stefano and Ibba, Giulia and Bartolucci, Stefano and others},
  title     = {DApps ecosystems: mapping the network structure of smart contract interactions},
  journal   = {EPJ Data Science},
  volume    = {13},
  pages     = {60},
  year      = {2024},
  doi       = {10.1140/epjds/s13688-024-00497-8},
  url       = {https://doi.org/10.1140/epjds/s13688-024-00497-8},
  received  = {2024-02-19},
  accepted  = {2024-09-02},
  published = {2024-09-27},
  note      = {Version of record 27 September 2024}
}

@inproceedings{winnertake_3,
author = {Yan, Kailun and Lu, Bo and Agrawal, Pranav and Li, Jiasun and Diao, Wenrui and Zhang, Xiaokuan},
title = {An Empirical Study on Cross-chain Transactions: Costs, Inconsistencies, and Activities},
year = {2025},
isbn = {9798400714108},
publisher = {Association for Computing Machinery},
address = {New York, NY, USA},
url = {https://doi.org/10.1145/3708821.3733878},
doi = {10.1145/3708821.3733878},
booktitle = {Proceedings of the 20th ACM Asia Conference on Computer and Communications Security},
pages = {939–954},
numpages = {16},
location = {
},
series = {ASIA CCS '25}
}

@InProceedings{solo_converge_1,
author="Gencer, Adem Efe
and Basu, Soumya
and Eyal, Ittay
and van Renesse, Robbert
and Sirer, Emin G{\"u}n",
editor="Meiklejohn, Sarah
and Sako, Kazue",
title="Decentralization in Bitcoin and Ethereum Networks",
booktitle="Financial Cryptography and Data Security",
year="2018",
publisher="Springer Berlin Heidelberg",
address="Berlin, Heidelberg",
pages="439--457",
isbn="978-3-662-58387-6"
}

@InProceedings{solo_converge_2,
author="Eyal, Ittay
and Sirer, Emin G{\"u}n",
editor="Christin, Nicolas
and Safavi-Naini, Reihaneh",
title="Majority Is Not Enough: Bitcoin Mining Is Vulnerable",
booktitle="Financial Cryptography and Data Security",
year="2014",
publisher="Springer Berlin Heidelberg",
address="Berlin, Heidelberg",
pages="436--454",
isbn="978-3-662-45472-5"
}

@inproceedings{solo_converge_3,
author = {Lewenberg, Yoad and Bachrach, Yoram and Sompolinsky, Yonatan and Zohar, Aviv and Rosenschein, Jeffrey S.},
title = {Bitcoin Mining Pools: A Cooperative Game Theoretic Analysis},
year = {2015},
isbn = {9781450334136},
publisher = {International Foundation for Autonomous Agents and Multiagent Systems},
address = {Richland, SC},
booktitle = {Proceedings of the 2015 International Conference on Autonomous Agents and Multiagent Systems},
pages = {919–927},
numpages = {9},
location = {Istanbul, Turkey},
series = {AAMAS '15}
}

@incollection{blockchain_upgrade_difficult1,
  author    = {Barrera, Cathy and Hurder, Stephanie},
  title     = {Blockchain Upgrade as a Coordination Game},
  booktitle = {Cryptoassets: Legal, Regulatory, and Monetary Perspectives},
  publisher = {Cambridge University Press},
  year      = {2019}
}

@article{blockchain_updgrade_difficult2,
  author  = {Zhu, G. and He, D. and An, H. and others},
  title   = {The governance technology for blockchain systems: a survey},
  journal = {Frontiers of Computer Science},
  year    = {2024},
  volume  = {18},
  pages   = {182813},
  doi     = {10.1007/s11704-023-3113-x},
  url     = {https://doi.org/10.1007/s11704-023-3113-x}
}

@inproceedings {BFT2,
	author = {Miguel Castro and Barbara Liskov},
	title = {Practical Byzantine Fault Tolerance},
	booktitle = {3rd Symposium on Operating Systems Design and Implementation (OSDI 99)},
	year = {1999},
	address = {New Orleans, LA},
	url = {https://www.usenix.org/conference/osdi-99/practical-byzantine-fault-tolerance},
	publisher = {USENIX Association},
	month = feb
}

@inproceedings{pos_safe_1,
author = {Kiayias, Aggelos and Russell, Alexander and David, Bernardo and Oliynykov, Roman},
year = {2017},
month = {07},
pages = {357-388},
title = {Ouroboros: A Provably Secure Proof-of-Stake Blockchain Protocol},
isbn = {978-3-319-63687-0},
doi = {10.1007/978-3-319-63688-7_12}
}

@article{pos_safe_2,
  title={Snow White: Provably Secure Proofs of Stake},
  author={Iddo Bentov and Rafael Pass and Elaine Shi},
  journal={IACR Cryptol. ePrint Arch.},
  year={2016},
  volume={2016},
  pages={919},
  url={https://api.semanticscholar.org/CorpusID:16970691}
}

@inproceedings{pos_safe_3,
author = {Gilad, Yossi and Hemo, Rotem and Micali, Silvio and Vlachos, Georgios and Zeldovich, Nickolai},
title = {Algorand: Scaling Byzantine Agreements for Cryptocurrencies},
year = {2017},
isbn = {9781450350853},
publisher = {Association for Computing Machinery},
address = {New York, NY, USA},
url = {https://doi.org/10.1145/3132747.3132757},
doi = {10.1145/3132747.3132757},
booktitle = {Proceedings of the 26th Symposium on Operating Systems Principles},
pages = {51–68},
numpages = {18},
location = {Shanghai, China},
series = {SOSP '17}
}

@article{rollup_layer2,
title = {A survey of Layer-two blockchain protocols},
journal = {Journal of Network and Computer Applications},
volume = {209},
pages = {103539},
year = {2023},
issn = {1084-8045},
doi = {https://doi.org/10.1016/j.jnca.2022.103539},
url = {https://www.sciencedirect.com/science/article/pii/S1084804522001801},
author = {Ankit Gangwal and Haripriya Ravali Gangavalli and Apoorva Thirupathi}
}

@article{rollup_layer22,
author = {Tremblay Thibault, Louis and Sarry, Tom and Senhaji Hafid, Abdelhakim},
year = {2022},
month = {08},
pages = {},
title = {Blockchain Scaling Using Rollups: A Comprehensive Survey},
volume = {PP},
journal = {IEEE Access},
doi = {10.1109/ACCESS.2022.3200051}
}

@article{hack1,
title = {The destabilising effects of cryptocurrency cybercriminality},
journal = {Economics Letters},
volume = {191},
pages = {108741},
year = {2020},
issn = {0165-1765},
doi = {https://doi.org/10.1016/j.econlet.2019.108741},
url = {https://www.sciencedirect.com/science/article/pii/S0165176519303714},
author = {Shaen Corbet and Douglas J. Cumming and Brian M. Lucey and Maurice Peat and Samuel A. Vigne}
}

@techreport{hack2,
  author      = {Azar, Pablo D. and Olivas, Sergio and Sinha, Nish D.},
  title       = {The Price of Processing: Information Frictions and Market Efficiency in DeFi},
  institution = {Federal Reserve Bank of New York},
  type        = {Staff Report},
  number      = {1153},
  month       = apr,
  year        = {2025},
  doi         = {10.59576/sr.1153},
  url         = {https://doi.org/10.59576/sr.1153}
}

@article{luna_death,
title = {Anatomy of a Stablecoin’s failure: The Terra-Luna case},
journal = {Finance Research Letters},
volume = {51},
pages = {103358},
year = {2023},
issn = {1544-6123},
doi = {https://doi.org/10.1016/j.frl.2022.103358},
url = {https://www.sciencedirect.com/science/article/pii/S1544612322005359},
author = {Antonio Briola and David Vidal-Tomás and Yuanrong Wang and Tomaso Aste}
}

@techreport{chainlink2,
  title       = {Chainlink 2.0: Next Steps in the Evolution of Decentralized Oracle Networks},
  author      = {Breidenbach, Lorenz and Cachin, Christian and Chan, Benedict and Coventry, Alex and Ellis, Steve and Juels, Ari and Koushanfar, Farinaz and Miller, Andrew and Magauran, Brendan and Moroz, Daniel and Nazarov, Sergey and Topliceanu, Alexandru and Tramer, Florian and Zhang, Fan},
  year        = {2021},
  month       = apr,
  note        = {v1.0},
  institution = {Chainlink Labs},
  url         = {https://research.chain.link/whitepaper-v2.pdf}
}

@article{tx_friction,
author = {Shampanier, Kristina and Mazar, Nina and Ariely, Dan},
title = {Zero as a Special Price: The True Value of Free Products},
journal = {Marketing Science},
volume = {26},
number = {6},
pages = {742-757},
year = {2007},
doi = {10.1287/mksc.1060.0254},

URL = { 
    
        https://doi.org/10.1287/mksc.1060.0254
    
    

},
eprint = { 
    
        https://doi.org/10.1287/mksc.1060.0254
    
    

}
}

@inproceedings {model_demand_1,
author = {Myeongjae Jeon and Shivaram Venkataraman and Amar Phanishayee and Junjie Qian and Wencong Xiao and Fan Yang},
title = {Analysis of {Large-Scale} {Multi-Tenant} {GPU} Clusters for {DNN} Training Workloads},
booktitle = {2019 USENIX Annual Technical Conference (USENIX ATC 19)},
year = {2019},
isbn = {978-1-939133-03-8},
address = {Renton, WA},
pages = {947--960},
url = {https://www.usenix.org/conference/atc19/presentation/jeon},
publisher = {USENIX Association},
month = jul
}

@inproceedings{market_demand_2,
author = {Hu, Qinghao and Sun, Peng and Yan, Shengen and Wen, Yonggang and Zhang, Tianwei},
title = {Characterization and prediction of deep learning workloads in large-scale GPU datacenters},
year = {2021},
isbn = {9781450384421},
publisher = {Association for Computing Machinery},
address = {New York, NY, USA},
url = {https://doi.org/10.1145/3458817.3476223},
doi = {10.1145/3458817.3476223},
booktitle = {Proceedings of the International Conference for High Performance Computing, Networking, Storage and Analysis},
articleno = {104},
numpages = {15},
location = {St. Louis, Missouri},
series = {SC '21}
}

@inproceedings{off_chain_upgrade_pattern1,
author = {Samuel, Justin and Mathewson, Nick and Cappos, Justin and Dingledine, Roger},
title = {Survivable key compromise in software update systems},
year = {2010},
isbn = {9781450302456},
publisher = {Association for Computing Machinery},
address = {New York, NY, USA},
url = {https://doi.org/10.1145/1866307.1866315},
doi = {10.1145/1866307.1866315},
booktitle = {Proceedings of the 17th ACM Conference on Computer and Communications Security},
pages = {61–72},
numpages = {12},
location = {Chicago, Illinois, USA},
series = {CCS '10}
}

@inproceedings{off_chain_upgrade_pattern2_1,
author = {Bodell III, William E and Meisami, Sajad and Duan, Yue},
title = {Proxy hunting: understanding and characterizing proxy-based upgradeable smart contracts in blockchains},
year = {2023},
isbn = {978-1-939133-37-3},
publisher = {USENIX Association},
address = {USA},
booktitle = {Proceedings of the 32nd USENIX Conference on Security Symposium},
articleno = {103},
numpages = {18},
location = {Anaheim, CA, USA},
series = {SEC '23}
}

@InProceedings{offchain_multisig,
author="Boneh, Dan
and Drijvers, Manu
and Neven, Gregory",
editor="Peyrin, Thomas
and Galbraith, Steven",
title="Compact Multi-signatures for Smaller Blockchains",
booktitle="Advances in Cryptology -- ASIACRYPT 2018",
year="2018",
publisher="Springer International Publishing",
address="Cham",
pages="435--464",
isbn="978-3-030-03329-3"
}

@article{Dao_voting_contract,
  title={Understanding Blockchain Governance: Analyzing Decentralized Voting to Amend DeFi Smart Contracts},
  author={Johnnatan Messias and Vabuk Pahari and Balakrishnan Chandrasekaran and Krishna P. Gummadi and Patrick Loiseau},
  journal={ArXiv},
  year={2023},
  volume={abs/2305.17655},
  url={https://api.semanticscholar.org/CorpusID:258959534}
}

@article{Dao_voting_contract_2,
title = {Analyzing voting power in decentralized governance: Who controls DAOs?},
journal = {Blockchain: Research and Applications},
volume = {5},
number = {3},
pages = {100208},
year = {2024},
issn = {2096-7209},
doi = {https://doi.org/10.1016/j.bcra.2024.100208},
url = {https://www.sciencedirect.com/science/article/pii/S2096720924000216},
author = {Robin Fritsch and Marino Müller and Roger Wattenhofer}
}

@misc{megatron_lm,
  author        = {Shoeybi, Mohammad and Patwary, Mostofa and Puri, Raul and LeGresley, Patrick and Casper, Jared and Catanzaro, Bryan},
  title         = {Megatron-LM: Training Multi-Billion Parameter Language Models Using Model Parallelism},
  year          = {2019},
  eprint        = {1909.08053},
  archivePrefix = {arXiv},
  primaryClass  = {cs.CL},
  url           = {https://arxiv.org/abs/1909.08053}
}

@misc{zero_rajbhandari2020,
  author        = {Rajbhandari, Samyam and Rasley, Jeff and Ruwase, Olatunji and He, Yuxiong},
  title         = {ZeRO: Memory Optimization Toward Training Trillion Parameter Models},
  year          = {2020},
  eprint        = {1910.02054},
  archivePrefix = {arXiv},
  primaryClass  = {cs.LG},
  url           = {https://arxiv.org/abs/1910.02054}
}

@misc{gpipe,
  author        = {Huang, Yanping and Cheng, Youlong and Bapna, Ankur and Firat, Orhan and Chen, Dehao and Chen, Mia and Lee, HyoukJoong and Ngiam, Jiquan and Le, Quoc V. and Wu, Yonghui},
  title         = {GPipe: Efficient Training of Giant Neural Networks using Pipeline Parallelism},
  year          = {2018},
  eprint        = {1811.06965},
  archivePrefix = {arXiv},
  primaryClass  = {cs.DC},
  url           = {https://arxiv.org/abs/1811.06965}
}

@inproceedings{fedavg2017,
  author    = {McMahan, Brendan and Moore, Eider and Ramage, Daniel and Hampson, Seth and Ag{\"u}era y Arcas, Blaise},
  title     = {Communication-Efficient Learning of Deep Networks from Decentralized Data},
  booktitle = {Proceedings of the 20th International Conference on Artificial Intelligence and Statistics},
  series    = {Proceedings of Machine Learning Research},
  volume    = {54},
  pages     = {1273--1282},
  year      = {2017}
}

@inproceedings{secureagg_ccs2017,
  author    = {Bonawitz, Keith and Ivanov, Vladimir and Kreuter, Ben and Marcedone, Antonio and McMahan, H. Brendan and Patel, Sarvar and Ramage, Daniel and Segal, Aaron and Seth, Karn},
  title     = {Practical Secure Aggregation for Privacy-Preserving Machine Learning},
  booktitle = {Proceedings of the 2017 ACM SIGSAC Conference on Computer and Communications Security},
  pages     = {1175--1191},
  year      = {2017},
  doi       = {10.1145/3133956.3133982}
}

@inproceedings{ckks2017,
  author    = {Cheon, Jung Hee and Kim, Andrey and Kim, Miran and Song, Yongsoo},
  title     = {Homomorphic Encryption for Arithmetic of Approximate Numbers},
  booktitle = {Advances in Cryptology -- ASIACRYPT 2017},
  publisher = {Springer},
  year      = {2017},
  pages     = {409--437},
  doi       = {10.1007/978-3-319-70694-8_15}
}

@inproceedings{abadi2016dp,
  author    = {Abadi, Mart{\'i}n and Chu, Andy and Goodfellow, Ian and McMahan, H. Brendan and Mironov, Ilya and Talwar, Kunal and Zhang, Li},
  title     = {Deep Learning with Differential Privacy},
  booktitle = {Proceedings of the 2016 ACM SIGSAC Conference on Computer and Communications Security},
  pages     = {308--318},
  year      = {2016},
  doi       = {10.1145/2976749.2978318}
}

@misc{pot_repo,
  title        = {Proof of Training (PoT) Artifact Repository},
  howpublished = {\url{https://github.com/P-HOW/proof-of-training}},
  note         = {Accessed: 2026-03-06}
}

@misc{eosio_consensus_doc,
  title        = {EOSIO Consensus Protocol (Documentation)},
  howpublished = {\url{https://github.com/EOSIO/welcome/blob/master/docs/60_protocol-guides/10_consensus_protocol.md}},
  note         = {Accessed: 2026-03-07}
}

@misc{eosio_dpos_study,
  title        = {An Empirical Study of EOSIO (DPoS) and Its Governance},
  howpublished = {\url{https://arxiv.org/abs/2211.05949}},
  note         = {arXiv:2211.05949, Accessed: 2026-03-07}
}

@misc{bnb_bep131,
  title        = {BEP-131: BNB Smart Chain Validator Set and Epoch Parameters (Specification)},
  howpublished = {\url{https://github.com/bnb-chain/BEPs/blob/master/BEP131.md}},
  note         = {Accessed: 2026-03-07}
}

@misc{bnb_security_case_study,
  title        = {A Case Study of BNB Smart Chain},
  howpublished = {\url{https://www.usenix.org/system/files/usenixsecurity25-li-rujia.pdf}},
  note         = {USENIX Security, Accessed: 2026-03-07}
}

@misc{tron_sr_doc,
  title        = {TRON Developer Documentation: Super Representatives},
  howpublished = {\url{https://developers.tron.network/docs/super-representatives}},
  note         = {Accessed: 2026-03-07}
}

@misc{tron_committee_study,
  title        = {An Analysis of TRON Committee-Based Block Production},
  howpublished = {\url{https://arxiv.org/html/2509.16292v1}},
  note         = {Accessed: 2026-03-07}
}

@misc{quorum_ibft_doc,
  title        = {Quorum IBFT (Istanbul Byzantine Fault Tolerance) Documentation},
  howpublished = {\url{https://github.com/ConsenSys/quorum-ibft}},
  note         = {Accessed: 2026-03-07}
}

@misc{ibft_analysis,
  title        = {On the Byzantine Fault Tolerance of IBFT},
  howpublished = {\url{https://arxiv.org/abs/1901.07160}},
  note         = {arXiv:1901.07160, Accessed: 2026-03-07}
}

@misc{fabric_orderer_plan,
  title        = {Hyperledger Fabric Deployment Guide: Ordering Service Planning},
  howpublished = {\url{https://ethan-li-fabric.readthedocs.io/en/latest/deployorderer/ordererplan.html}},
  note         = {Accessed: 2026-03-07}
}

@misc{fabric_arch_paper,
  title        = {Hyperledger Fabric: A Distributed Operating System for Permissioned Blockchains},
  howpublished = {\url{https://arxiv.org/abs/1801.10228}},
  note         = {arXiv:1801.10228, Accessed: 2026-03-07}
}

\end{document}